\documentclass[11pt,a4paper]{article}
\pdfoutput=1 

\usepackage{jheppub}
\graphicspath{{plots/}} 

\usepackage{bbm}

\newcommand{\Tr}{\mathrm{Tr}}

\newcommand{\order}[1]{\mathcal{O}\left(#1\right)}
\newcommand{\m}{\phantom{$-$}}

\newcommand{\new}[1]{{#1}}

\makeatletter \@addtoreset{equation}{section} \makeatother

\title{An Improved Single-Plaquette Gauge Action}

\author[a,e]{D.\ Banerjee,} \author[a,b]{M.\ B\"ogli,} \author[c]{K.\ Holland,}
\author[a]{F.\ Niedermayer,} \author[d]{M.\ Pepe,} \author[a]{U.\ Wenger,}
\author[a]{and U.-J.\ Wiese}

\affiliation[a]{Albert Einstein Center for Fundamental Physics \\
  Institute for Theoretical Physics, University of Bern, Sidlerstr.\ 5, 3012
  Bern, Switzerland}
\affiliation[b]{Department of Physics \\
  Chung-Yuan Christian University (CYCU), Chung-Li 32023, Taiwan}
\affiliation[c]{Physics Department \\
  University of the Pacific, 3601 Pacific Avenue, Stockton, CA 95211, USA}
\affiliation[d]{INFN, Istituto Nazionale di Fisica Nucleare \\
  Sezione di Milano-Bicocca, Edificio U2, Piazza della Scienza 3 - 20126
  Milano, Italy}
\affiliation[e]{NIC, DESY, Platanenallee 6, 15738, Zeuthen, Germany}

\emailAdd{debasish.banerjee@desy.de}
\emailAdd{michael.boegli@gmail.com}
\emailAdd{kholland@pacific.edu}
\emailAdd{niedermayer@itp.unibe.ch}
\emailAdd{michele.pepe@mib.infn.it}
\emailAdd{wenger@itp.unibe.ch}
\emailAdd{wiese@itp.unibe.ch}

\abstract{We describe and test a nonperturbatively improved single-plaquette
  lattice action for 4-d SU(2) and SU(3) pure gauge theory, which suppresses
  large fluctuations of the plaquette, without requiring the
  naive continuum limit for smooth fields. 
  We tune the action parameters based on
  torelon masses in moderate cubic physical volumes, and investigate the size
  of cut-off effects in other physical quantities, including torelon masses in
  asymmetric spatial volumes, the static quark potential, and gradient flow
  observables. In 2-d O($N$) models similarly constructed nearest-neighbor
  actions have led to a drastic reduction of cut-off effects, down to the
  permille level, in a wide variety of physical quantities. In the gauge
  theories, we find significant reduction of lattice artifacts, and for some
  observables, the coarsest lattice result is very close to the continuum
  value. We estimate an improvement factor of 40 compared to using the 
  Wilson gauge action to achieve the same statistical accuracy and 
  suppression of cut-off effects. The simplicity of the gauge action 
  makes it amenable for dynamical fermion simulations.}

\begin{document}

\maketitle

\section{Introduction}
\label{sec:intro}

Cut-off effects are a major source of systematic errors in lattice QCD
calculations. Improved lattice actions are valuable for obtaining reliable
continuum results, but usually imply an increased numerical effort.  The
Symanzik improvement program is a systematic method to eliminate cut-off
effects, order by order in the lattice spacing $a$, by including additional
operators in the lattice action, beyond the standard plaquette term
\cite{Sym83,Sym83a,Lue85,Lue85a}. These operators, which have a larger
space-time extent than the standard term, lead to greater numerical cost in
Monte Carlo simulations. The coefficients of the additional operators can be
fixed either perturbatively, by expanding the lattice operators in continuum
operators of increasing dimension, or nonperturbatively by adjusting them to
satisfy continuum physics constraints.  A more radical improvement strategy
underlies the perfect action approach, which attempts to eliminate cut-off
effects to all orders of $a$, at least at the classical level
\cite{Hasenfratz:1993sp,Bietenholz:1995cy,Niedermayer:2000yx,
Hasenfratz:2002rp,Gattringer:2003qx}.
The classically perfect fixed point action, which is located on the critical
surface at the end of a renormalized trajectory, is very complicated. Still,
it can be parametrized to high accuracy with a relatively large number of
terms, which is thus costly.  The parametrized fixed point action is then used
for numerical simulations of the quantum theory away from the critical
surface. This has led to a substantial reduction of cut-off effects in a
variety of asymptotically free field theories, ranging from the 2-d O(3) model
to QCD. A different approach uses mixed fundamental-adjoint actions to reduce
cut-off effects without extending the space-time extent of the operators
beyond a single plaquette, and thus with only a moderate increase of the
computational cost
\cite{Ardill:1982nk,Blum:1994xb,Heller:1995bz,Hasenbusch:2004yq}.
In \cite{Bietenholz:2005rd,Fukaya:2005cw} the cut-off effects 
were studied using a modified single-plaquette gauge action proposed 
in \cite{Luscher:1998du} for theoretical purposes. 

2-d O(N) models share several features, including asymptotic freedom and a
nonperturbatively generated mass gap, with 4-d non-Abelian gauge theories.
Hence, they serve as a good testing ground for lattice improvement studies.
Interestingly, in contrast to what Symanzik improvement suggests, the cut-off
effects in the 2-d O(3) model appeared to be ${\cal O}(a)$ instead of ${\cal
  O}(a^2)$ \cite{Hasenbusch:2001ht,Knechtli:2005jh}. A careful analysis of
this apparent contradiction verified the Symanzik ${\cal O}(a^2)$ prediction,
but showed that easily accessible lattice spacings are affected by large
logarithmic corrections, which mimic ${\cal O}(a)$ behavior
\cite{Balog:2009yj,Balog:2009np}.  Recently, different lattice actions have
been studied with the goal to improve the cut-off effects
\cite{Bogli:2011aa}. In addition to the standard action, this study used a
topological action \cite{Bietenholz:2010xg}, which constrains the maximal
angle between neighboring spins and is therefore invariant under small field
deformations. Although it does not have the correct naive continuum limit and
it violates the Schwarz inequality between the action and the topological 
charge, it still yields the correct quantum continuum limit
\cite{Bogli:2011aa}. Combining this action with the standard action one gets
an improved constrained action, which eliminates the lattice spacing effects
almost entirely. Using this improved action, it was possible to study the
$\theta$-vacuum angle in the 2-d O(3) model, which turned out to be a relevant
parameter of the theory that does not get renormalized non-perturbatively. For
the first time, this numerically confirmed the conjectured exact S-matrix
results at $\theta = \pi$ \cite{Balogpriv} beyond any reasonable doubt. This
also confirmed the existence of a conformal fixed point at $\theta = \pi$,
where the model reduces to the WZNW model at low energies.  This study has
also been a basis for further investigations to demonstrate walking near the
conformal fixed point close to $\theta \approx \pi$ \cite{deForcrand:2012se}.
The essential features of walking technicolor models are shared by this toy
model and can be accurately investigated by numerical simulations. Optimized
lattice actions have also been studied intensively in \cite{Balog:2012db} for
2-d O($N$) models, where it has been shown that cut-off effects can be reduced
to the per mille level. A topological lattice action has also been used 
in a recent study of 4-d U(1) gauge theory, to demonstrate that the correct 
continuum limit is obtained, to examine the effect of the lattice action 
on monopole condensation in the confined phase, and to test a method to 
measure the free energy \cite{Akerlund:2015zha}.

In this work, we apply a similar strategy to gauge theories. Our approach is
different from Symanzik's improvement program \cite{Sym83}, where one adds
operators with higher dimensions to the standard action to eliminate the
leading cut-off effects.  The Symanzik improved action is perturbative, even
if the coefficients are determined non-perturbatively.  The experience with
the O($N$) non-linear sigma model suggests that at moderate lattice spacing
used in the numerical simulations the main cause of the cut-off effects are
the large local fluctuations of the action density.  A truly non-perturbative
action, like the topological action or the constrained action with negative
$\beta$ performs surprisingly well in that case.  With one extra parameter
(and a ``cheap'' action) one can reach a strong suppression of cut-off
effects.
 
For the SU(2) and SU(3) pure gauge theory here we study a slight
modification of the constrained action.  We found that the improved
action decreases the cut-off effects of many quantities including
torelon masses on asymmetric lattices, the static potential, and
observables related to the gradient flow of the gauge fields.  The
simplicity of the gauge action makes it easy to embed it in simulations 
of gauge theories with dynamical fermions at little extra cost, with 
the possible gain of reduced lattice artifacts.

\new{ Our approach differs from those used in \cite{Hasenbusch:2004yq}
and \cite{Bietenholz:2005rd,Fukaya:2005cw} in two aspects. 
Our set of single-plaquette actions considered is broader --
it includes the possibility $\beta\le 0$ (see below).
In other words, we do not restrict ourselves to actions
having a naive continuum limit, hence we can optimize the action on 
coarser lattices as well. 
Secondly, we use the torelon masses in small boxes to optimize the action 
-- these are much easier to measure than the string tension $\sigma$ 
or the deconfining temperature $T_c$.
}

This paper is organized as follows. In section \ref{sec:tuning} we discuss the
parametrization of the improved action, the procedure for optimization of the
action parameters, and some basic properties of torelon states. Section
\ref{sec:SU2} shows our simulation results for SU(2) gauge theory, where we
describe the tuning of the action parameters and the reduction of lattice
artifacts for torelons on asymmetric spatial volumes and for the static quark
potential. In section \ref{sec:SU3} we present similar findings for SU(3)
gauge theory, as well as a study of the cut-off dependence of observables
obtained from the gradient flow of the gauge fields. Section \ref{sec:costs}
details what algorithms we use and the numerical cost of Monte Carlo
simulations with this novel action. We finish with our conclusions in section
\ref{sec:conclusion}.

\section{Determination of the parameters of the action}
\label{sec:tuning}

Consider the constrained action for pure Yang-Mills theory with the action
density associated with the plaquette
\begin{equation}
  S_p =
  \begin{cases}
    \beta w \,, & \text{ for $w < \delta$} \,, \\
    \infty \,, & \text{ otherwise} \,.
  \end{cases}
  \label{splaq} 
\end{equation}
Here $w = 1- \frac{1}{N} \text{Tr}U_p$, where $U_p$ is the standard plaquette
matrix, and plaquette values larger than the constraint $\delta$ are
prohibited.  Keeping in mind that the gauge action could be used in Hybrid
Monte Carlo simulations, we have chosen a smooth version of the constrained
action with
\begin{equation}
  S_p = \beta w + \gamma w^q \,.
  \label{impaction} 
\end{equation}
For large power $q$ this has the same effect as the constrained action with
$\delta\approx \gamma^{-1/q}$.  In our simulations we used a fixed value of
the power, $q=10$.\footnote{Since here we used a standard Metropolis update,
  we could have chosen the constrained action as well.}  To reduce the cut-off
effects one can choose two appropriate physical quantities. One of them is
used to set the lattice spacing $a$, the other to estimate the size of the
cut-off effects at the given resolution. For the 2-d O($N$) spin model these
were \cite{Balog:2012db} the mass gap measured on a long strip with spatial
sizes $L$ and $2L$ (cf. step scaling function, \cite{Luscher:1991wu}).

\begin{figure}[tbp]
  \centering
  \includegraphics[width=.45\textwidth]{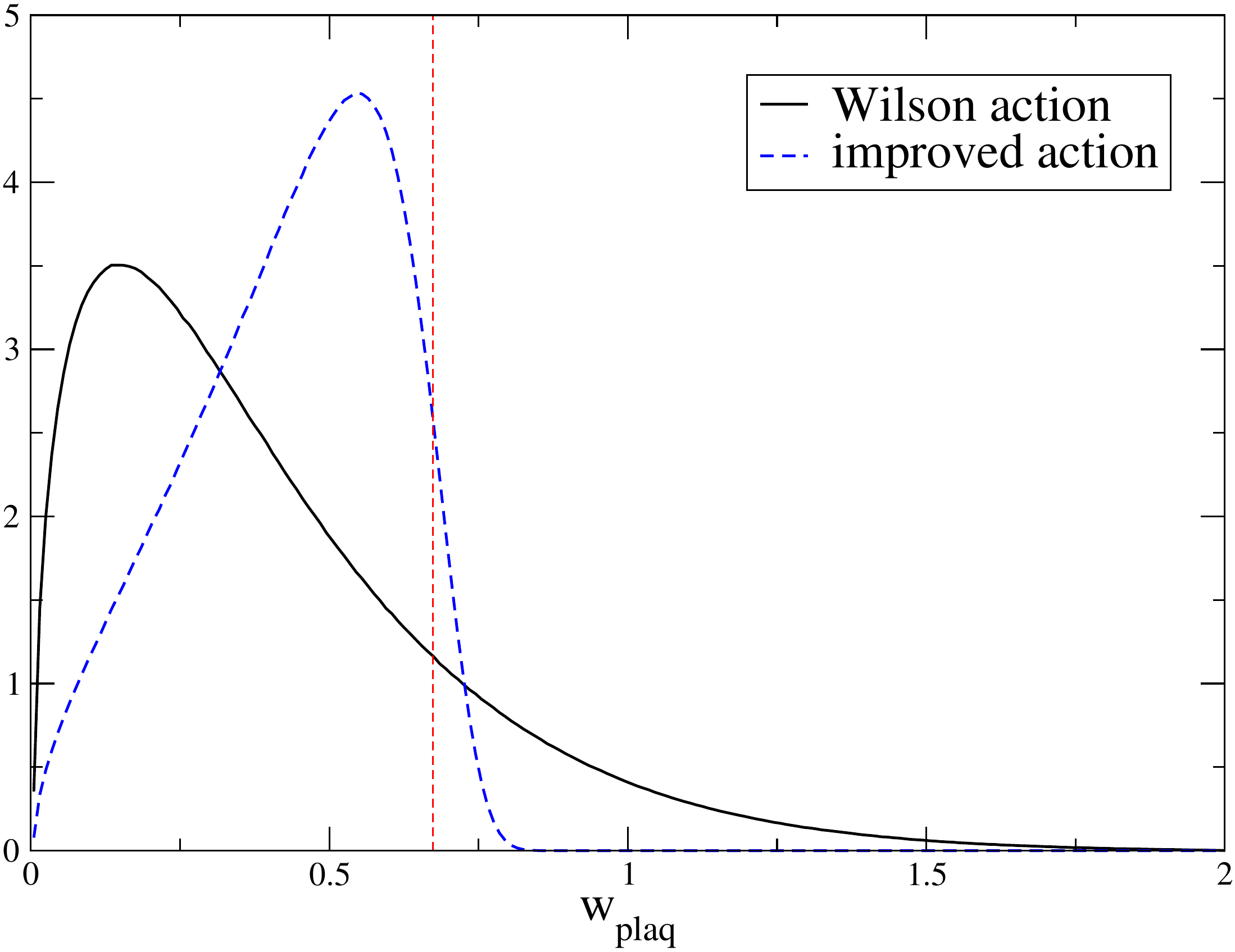}
  \hfill
  \includegraphics[width=.45\textwidth]{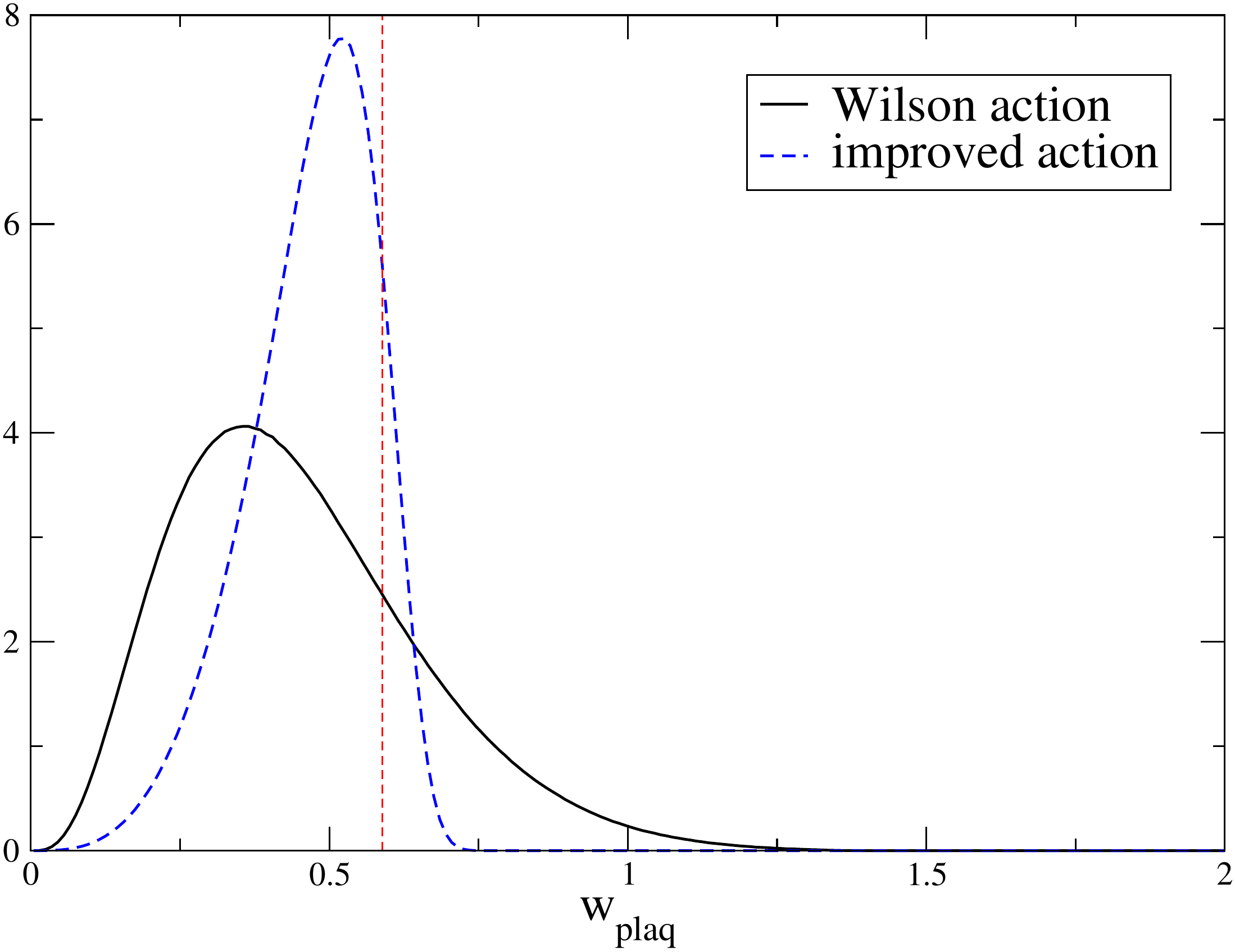}
  \caption{Histograms of the plaquette variable $w$ for the Wilson action and
    the improved action in the SU(2) and SU(3) case.  The vertical lines show
    the corresponding effective cut $\gamma^{-1/q}$. The distributions were
    generated from simulations on $(L'/a)^4 = 10^4$ volumes, where the lattice
    spacing was set to its coarsest value, defined by $L/a=4$ and $m_{100}(L)L
    = 1.375$ and 1 for SU(2) and SU(3) respectively.}
  \label{histograms}
\end{figure}

For the gauge theory we considered the energy gap between the vacuum state and
states with given electric flux wrapping around the periodic spatial
directions \cite{tHooft}, in short the ``torelon masses''.  The two quantities
used for optimizing the action were the torelon masses $m_{100}(L)$ and
$m_{110}(L)$ in an $L^3$ spatial box, corresponding to fluxes wrapping around
along an axis and along a diagonal.  Our procedure was the following. 
Taking first a lattice of size
$L^3 \times L_t$ we determined a line $\beta=\beta(\gamma)$ along which
$u_{100}(L) \equiv m_{100}(L) L = u^\star$ is fixed.  (For SU(2) we took
$u^\star=1.375$, while for SU(3) $u^\star=1.0$.)  Note that $\beta(\gamma)$ is
a decreasing function, and at some $\gamma$ it becomes negative.  It is
important that we do not restrict ourselves to the $\beta > 0$
region.\footnote{The negative $\beta$ values are needed to compensate the
absence of coarse plaquettes. To reach the given lattice spacing one needs
to suppress simultaneously the very smooth plaquettes as well.}  The
distribution of the plaquette variable $w$ for the standard action and for the
improved action at $a T_c\sim 1/4$ is shown in Fig.~\ref{histograms}.

The optimization of the action is done by moving along the $\beta(\gamma)$
line to minimize the deviation of $u_{110}(L) \equiv m_{110}(L) L$ from its
continuum limit.  The latter was obtained by measuring the corresponding
torelon masses at the same physical point $u_{100}(L)=u^\star$ and finer
resolutions with the Wilson action and extrapolating to $a/L=0$.  This way one
obtains the pair of couplings ($\beta,\gamma$) which are optimal for this
resolution (and the given choice of observables).  One could proceed further
on lattices with larger $L/a$, similarly to the case of the O($N$) spin model.
Instead, we have chosen a less ambitious optimization: for finer resolutions
we kept the same $\gamma$ as obtained on the coarse lattice $L/a=4$, and tuned
only $\beta$ to obtain $u_{100}(L)=u^\star$ for $L/a=6$ and $8$.  Then using
these pairs of couplings ($\beta(a),\gamma$) we measured different quantities,
like torelon masses on spatial lattices of different shapes, the static
$q\overline{q}$ potential, and observables related to the gradient flow of the
gauge fields.

It is worth to discuss briefly our choice of the basic physical quantity used
for optimization.  The torelon is an excitation characterized by an electric
flux wrapping through the torus $L_1\times L_2\times L_3\times\infty$ in one
(or several) periodic spatial directions.  The Hilbert space of the transfer
matrix is split into sectors characterized by quantum numbers
$\mathbf{k}=(k_1,k_2,k_3)$, where $k_i=0,\ldots,N-1$ for SU($N$).  In
particular, the value $k_1$ describes the transformation property of the
corresponding state (the wave function in the strong-coupling basis) with
respect to the multiplication by $z=\exp(2\pi i/N)$ of links
$U_1(x_1,x_2,x_3,t)$ at some plane with a given fixed $x_1$.  Such a
state can be created from a state in the vacuum sector $\mathbf{k}=(0,0,0)$ by
multiplying it by a product of traces of Polyakov loops, or by the trace of a
single Polyakov type loop wrapping around $(k_1,k_2,k_3)$ times the spatial
volume.  We denote the trace of the Polyakov loop $\Phi_i$ in the direction
$i$ on time slice $t$ by\footnote{suppressing the coordinates in the
  transverse direction}
\begin{equation}
  \phi_{i}(t) = \Tr \,\Phi_{i}(t)\,, \quad i=1,2,3 \,,
\end{equation}
and
\begin{equation} \label{phik} \phi_\mathbf{k}(t) = \phi_1(t)^{k_1}
  \phi_2(t)^{k_2}\phi_3(t)^{k_3}\,,\quad k_i=0,\ldots,N-1 \,.
\end{equation}
With this the torelon mass $m_\mathbf{k}=m_\mathbf{k}(L_1,L_2,L_3)$ is
obtained from the exponential fall-off of the correlation function
\begin{equation}
  \langle \phi_\mathbf{k}(0) \phi_\mathbf{k}^\dagger(t)
  \rangle \sim A \exp(-m_\mathbf{k} t) \,.	
\end{equation}

The torelon mass (more precisely the energy difference between the lowest
state in the sector characterized by electric flux $\mathbf{k}$ and the vacuum
state) has a special dependence on the size and shape of the 3-volume.  For
small volumes $L\ll 1/T_c$ it is extremely small (it is given by a tunneling
through a high barrier \cite{vanBaal}). In this case the flux is completely
spread in the transverse direction. In a cubic 3-volume $L^3$ with
increasing $L$ the flux assumes a finite width (``flux tube'') while its
energy increases as $m_{100}(L)\sim\sigma L$, where $\sigma$ is the string
tension.  There is a relatively sharp transition between these two regimes,
and we have chosen our physical lattice sizes (i.e.~the value of $u^\star$) to
be roughly in this region.

For asymmetric volumes $m_{100}(L_1,L_2,L_3)$ increases with $L_1$, and
decreases with increasing transverse sizes $L_2$, $L_3$.  For $L_1=1/T_c$,
$L_2=L_3=\infty$ the system undergoes a phase transition.  We work here,
however, in volumes where all spatial sizes $L_i$ are of $\order{1/T_c}$,
hence the observables are smooth functions of $\beta$.

\section{The SU(2) case}
\label{sec:SU2}

\begin{figure}[tbp]
  \centering
  \includegraphics[width=.45\textwidth]{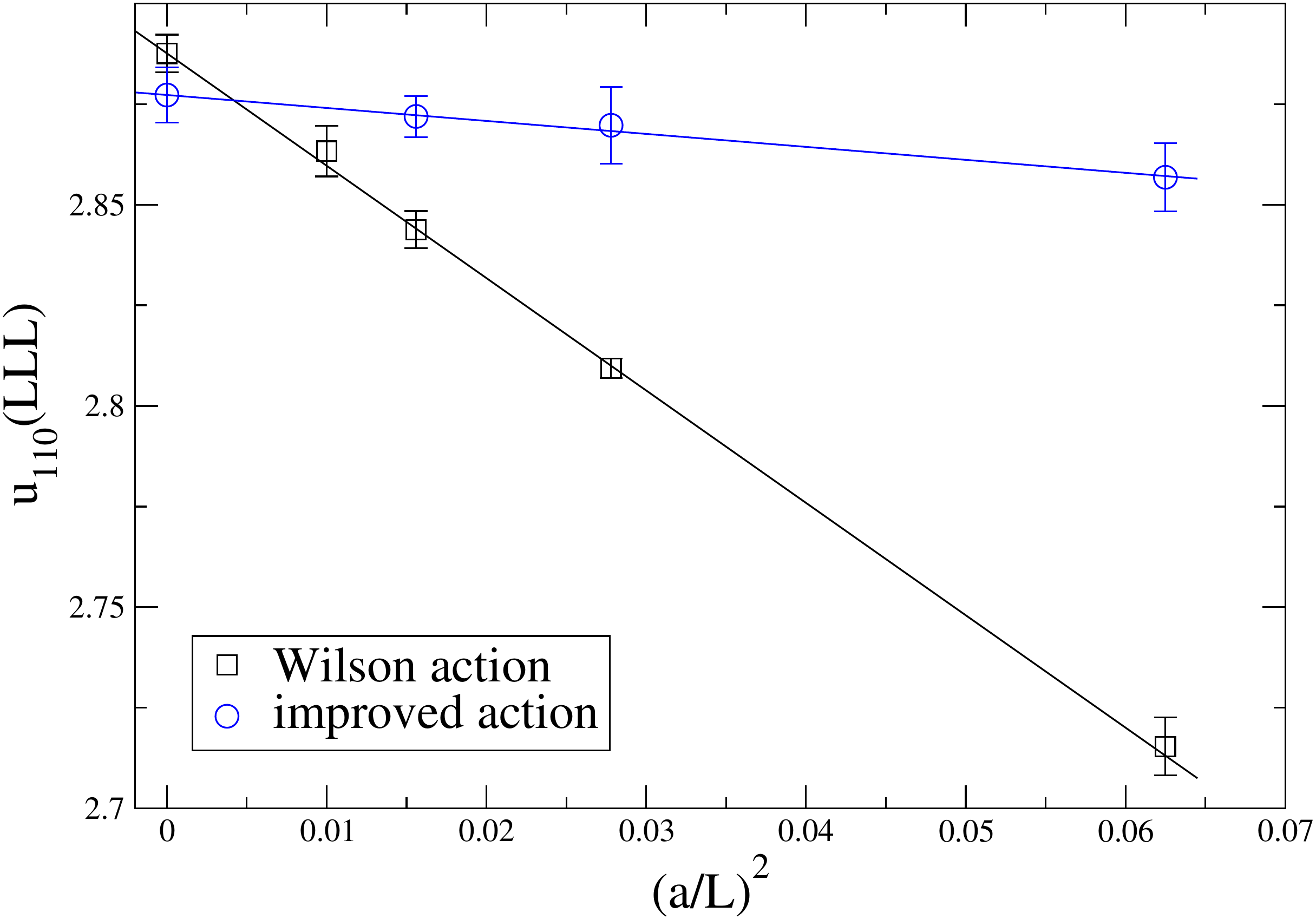}
  \caption{Cut-off effects of $u_{110}(L)$ on cubic lattices $L^3$ for the
    Wilson action and the improved action at $u_{100}(L)=1.375$ for SU(2).}
  \label{SU2_u110_LLL}
\end{figure}

On a cubic spatial volume $L^3$ we define the fixed physical volume via the
dimensionless combination $m_{100} L \equiv u_{100}(L) = u^\star = 1.375$
i.e.~the lattice size is measured in torelon mass units. At the chosen
$u^\star$ value we measured the diagonal torelon state $u_{110}(L)$ on cubic
spatial lattices of size $L/a=4,6,8,10$ using the Wilson action.  
The temporal extent $L_t$ was chosen to be either $10L$ or $20L$, the former 
corresponding to free boundary conditions in the time direction and the 
latter to periodic boundary conditions in the time direction. 
The advantage of free boundary conditions is that one can use a smaller
lattice volume, with the drawback that time-like correlators can only be
measured sufficiently far way from the ends. We used both setups to
cross-check that they give consistent determinations of the torelon mass. 
The extrapolation to the continuum limit using a linear fit in $a^2$ gave
$u_{110}(L)=2.888(5)$ (cf. figure~\ref{SU2_u110_LLL}).

To optimize the couplings in eq.~\eqref{impaction} we tried first the choice 
$q=2$ on a $4^3\times 80$ lattice, but a more significant improvement of 
the cut-off effect has been found by increasing the power, and for the rest 
of our simulations we took $q=10$.  Increasing $\gamma$ from the standard 
action case ($\gamma=0$) helps to decrease the cut-off effect for
$u_{110}(L)$. We went up with this until $\gamma=52$, corresponding to an 
effective cut $\delta=\gamma^{-1/q} = 0.67$ for $w$.  
At $L/a=4$ and $\gamma=52$ the condition
\begin{equation}
  u_{100}(L) \equiv m_{100}(L) L = u^\star
  \label{cond0} 
\end{equation}
yields $\beta=-2.4811$ (cf. table~\ref{SU2_par}).  Increasing $\gamma$ further
-- i.e.~decreasing the effective constraint $\delta$ -- lowers $\beta$, and the
action density becomes restricted practically to a narrow region for $w$
closely below $\delta$.  This would slow down significantly the effectiveness
of the Monte Carlo simulations.  We repeated the procedure for the improved 
action on $L/a = 6$ and $8$, holding $q=10$ and $\gamma=52$ fixed.

The $u_{110}$ torelon masses measured using the standard Wilson action and
separately the improved action are plotted in figure \ref{SU2_u110_LLL},
together with the extrapolations to the continuum limit. The couplings of the
actions are given in table~\ref{SU2_par}, where we include an extrapolation
from simulations to the chosen value of the physical point
$u_{100}(L)=u^\star=1.375$.  For the error propagation we used
$du_{110}(L)/du_{100}(L)\approx 1.8$, measured by repeating the
simulation\footnote{The derivative $\partial u_\mathbf{k}/\partial\beta$ can
  also be obtained at the given $\beta$ by measuring an appropriate
  correlation function. We used this method in a few cases.}  at slightly
different $\beta$.

In contrast to the situation with the O($N$) non-linear sigma model, for the
SU($N$) case we could not completely eliminate the cut-off effects for the
chosen pair of physical quantities, $u_{100}(L)$ and $u_{110}(L)$, on the
coarsest $L/a = 4$ lattice using the single-plaquette improved action with
only one tunable parameter. However, the improved action significantly reduces
the cut-off effect at $L/a = 4$ down to 1\%, compared to 6\% for the Wilson
action, and by $L/a=6$ the improved action result is essentially compatible
with the continuum value. For both actions, the lattice artifacts appear to be
${\cal O}(a^2)$ and we make our continuum extrapolations assuming quadratic
dependence on the lattice spacing. We see very good agreement between the
extrapolated values for the two lattice actions. We could make a more accurate
determination for the continuum value of $u_{110}(L)$ with a constrained fit
of Wilson and improved action data which demands that they have a common
continuum limit, but that would not serve our purpose here to check for
consistency between the two independent sets of simulations.

\begin{table}[htb]
  \centering
  \begin{tabular}{rlcllll}
    \phantom{9}$L/a$ & \phantom{9999}$\beta$ & $\gamma$ 
    & \phantom{9} $u_{100}(L)$ &  \phantom{9}$u_{110}(L)$ &
    \phantom{9}$u_{110}^\star(L)$ &  \phantom{9}$a$[fm] \\ 
    \hline
    4 &  \m2.29342  &  0. & 1.3746(8)  & 2.7148(71) & 2.7154(72) & 0.175\\
    6 &  \m2.42660  &  0. & 1.3754(4)  & 2.8100(23) & 2.8093(24) & 0.112\\
    8 &  \m2.51350  &  0. & 1.3756(7)  & 2.8450(44) & 2.8438(46) & 0.082\\
    10 &  \m2.58220  &  0. & 1.3769(13) & 2.8667(58) & 2.8633(63) & 0.064\\
    \hline
    4 &  $-2.4811$ & 52. & 1.3737(16) & 2.8544(81) & 2.8568(85) & \\
    6 &  $-1.3720$ & 52. & 1.3735(19) & 2.8670(89) & 2.8697(95) & \\
    8 &  $-0.7770$ & 52. & 1.3730(13) & 2.8683(46) & 2.8719(51) & \\
    \hline
  \end{tabular}
  \caption{Parameters of the SU(2) action and $u_{100}(L)$, $u_{110}(L)$ 
    on cubic lattices at different lattice spacings, for the Wilson 
    and the improved actions.
    The column $u_{110}^\star(L)$ is the value extrapolated to
    $u_{100}(L)=1.375$. For reference we include for the Wilson action
    the value of the lattice spacing as set via the string tension with
    $\sqrt{\sigma} = 420$~MeV \cite{Lucini:2005vg}.} 
  \label{SU2_par}
\end{table}

\subsection{Scaling Tests}
\label{sec:SU2_tests}

Once the parameters of the action have been set as described above, we can
proceed to examine the cut-off dependence of other physical quantities, to see
what improvement the new action delivers. We start with a discussion of
torelon masses measured on asymmetric spatial volumes.

\begin{figure}[tbp]
  \centering
  \includegraphics[width=.45\textwidth]{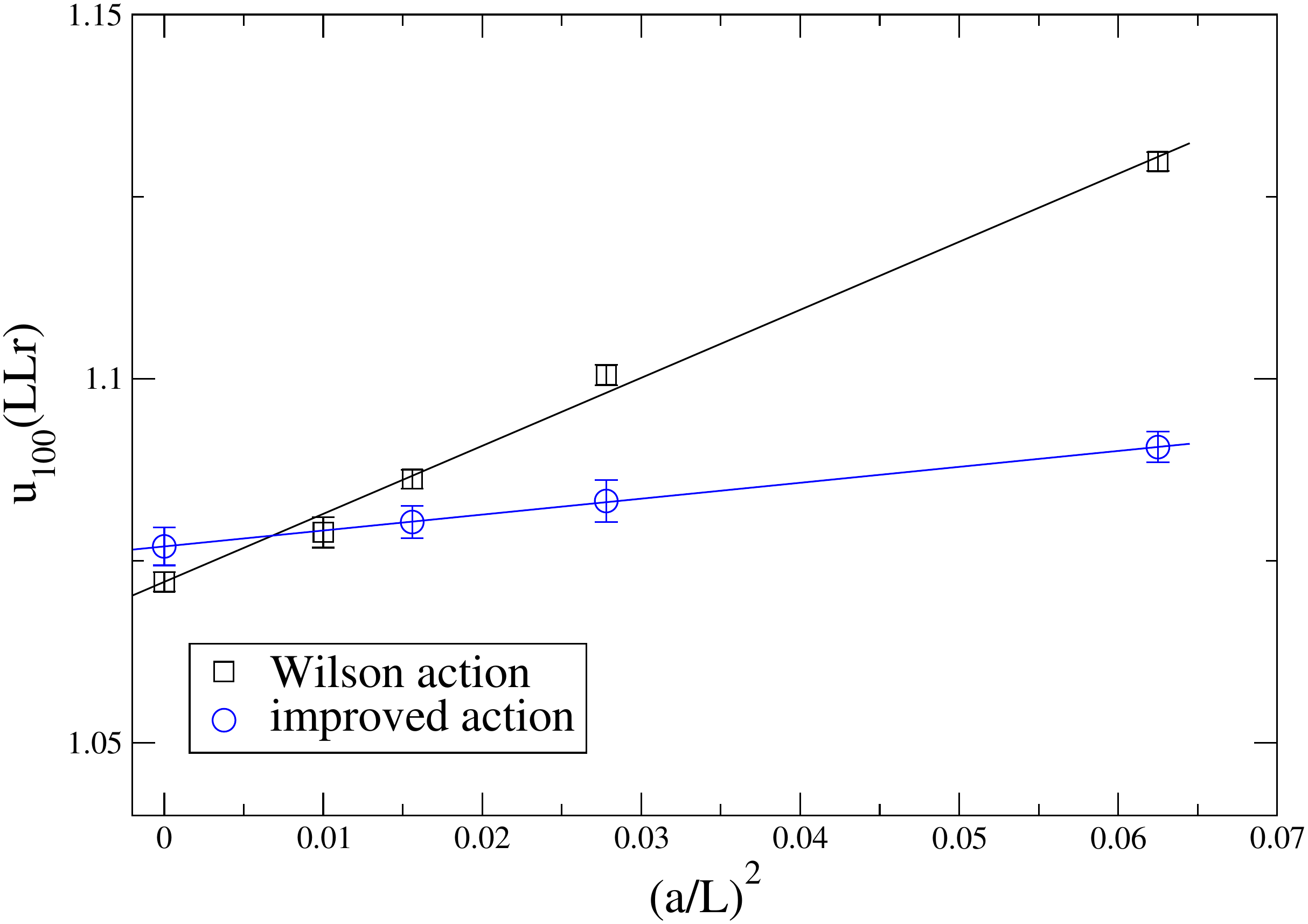}
  \hfill
  \includegraphics[width=.45\textwidth]{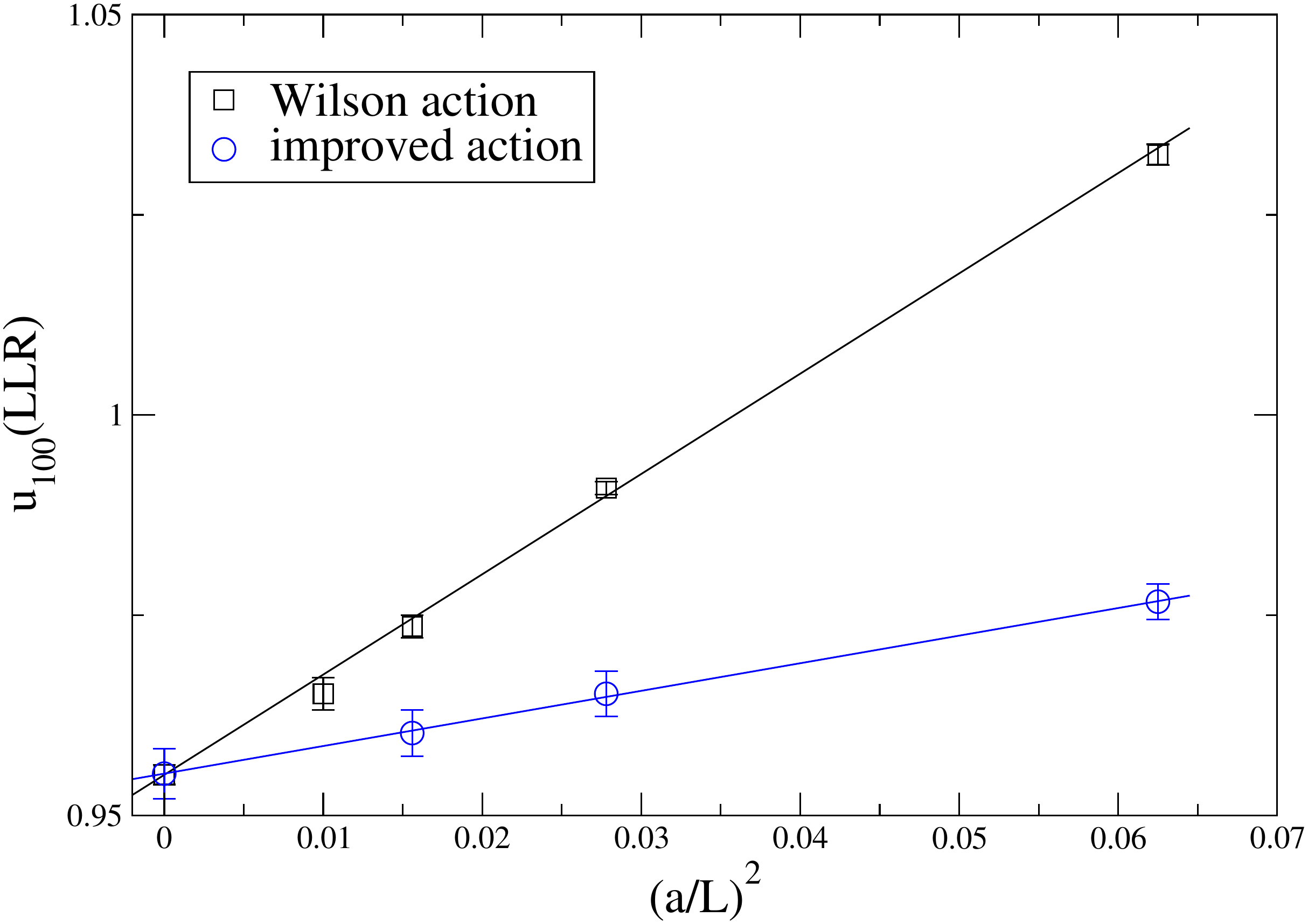}
  \caption{The $u_{100}$ torelon mass for SU(2) measured on spatial volumes
    (left) $(L,L, 3L/2)$ and (right) $(L,L, 2L)$.}
  \label{SU2_u100_LLx}
\end{figure}

\begin{figure}[tbp]
  \centering
  \includegraphics[width=.45\textwidth]{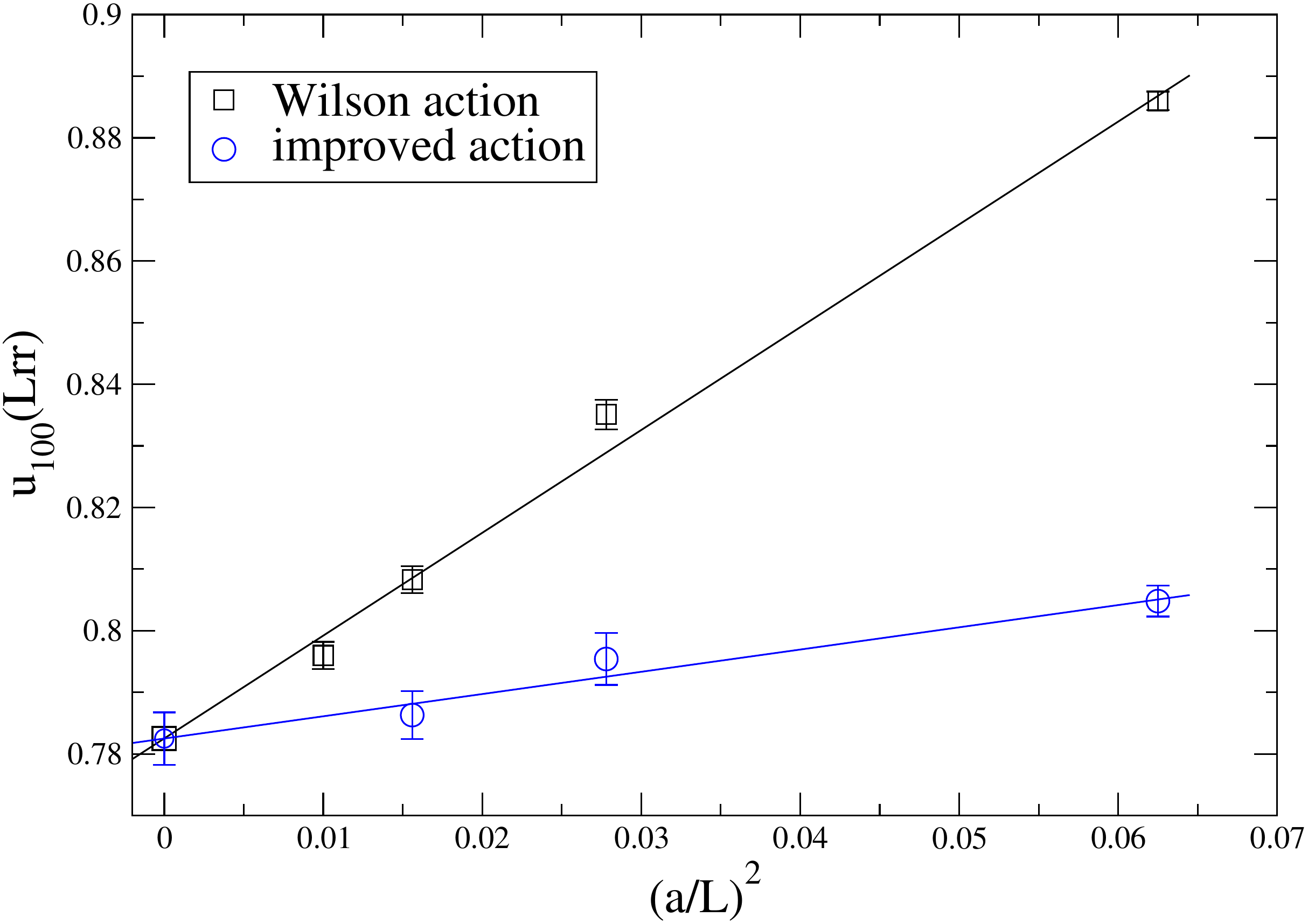}
  \hfill
  \includegraphics[width=.45\textwidth]{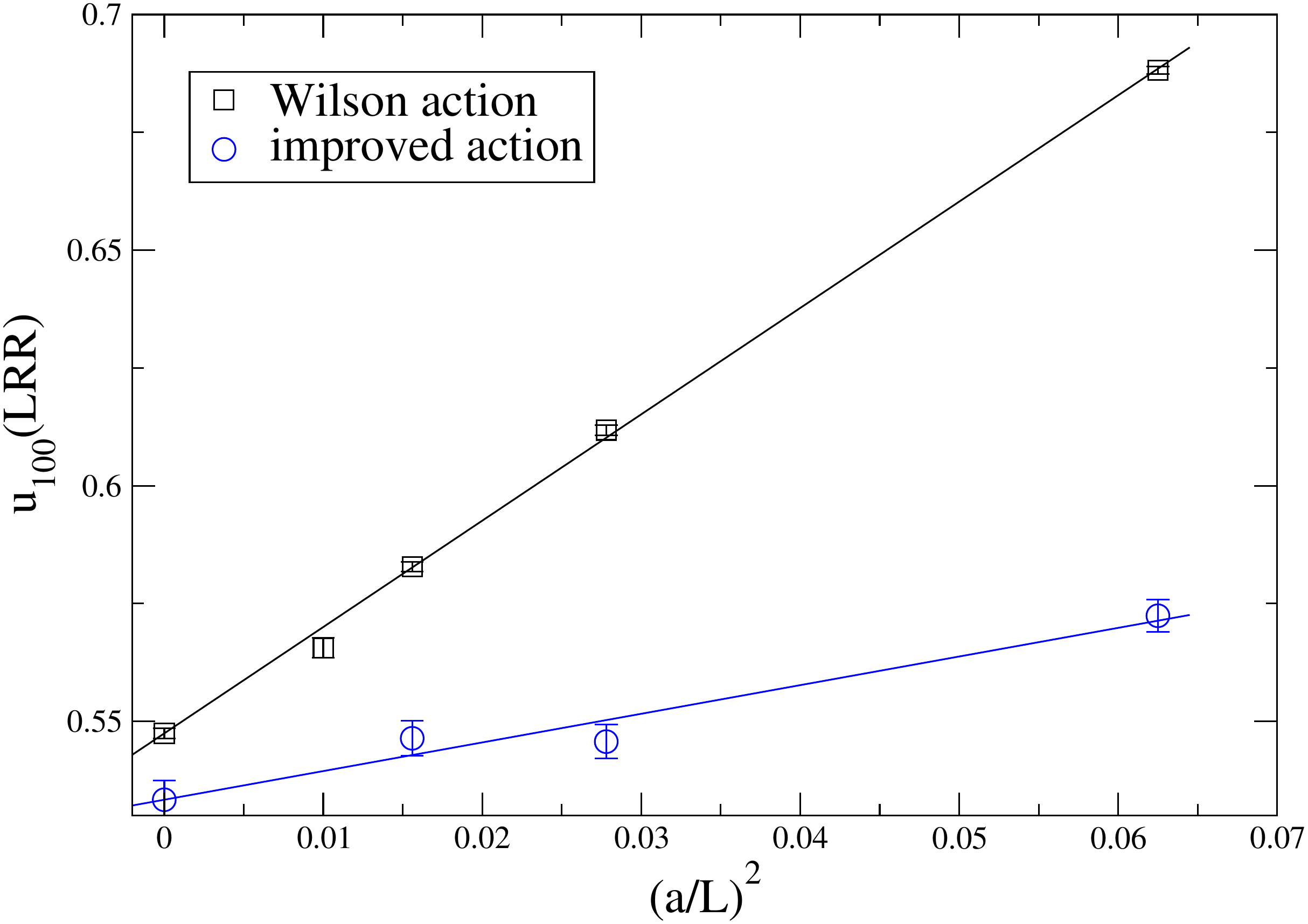}
  \caption{The $u_{100}$ torelon mass for SU(2) measured on spatial volumes
    (left) $(L,3L/2, 3L/2)$ and (right) $(L,2L, 2L)$.}
  \label{SU2_u100_Lxx}
\end{figure}

\begin{figure}[tbp]
  \centering
  \includegraphics[width=.45\textwidth]{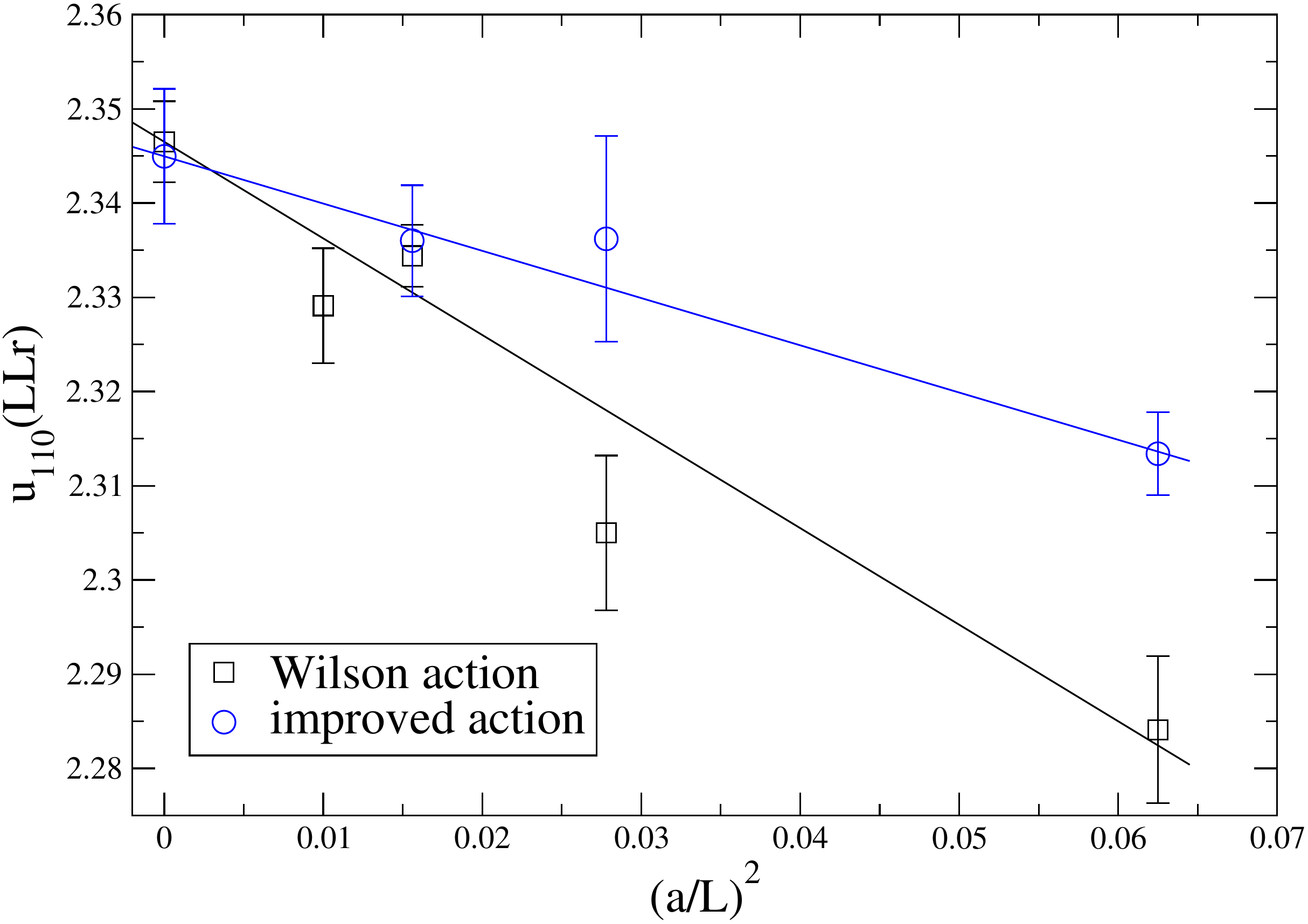}
  \hfill
  \includegraphics[width=.45\textwidth]{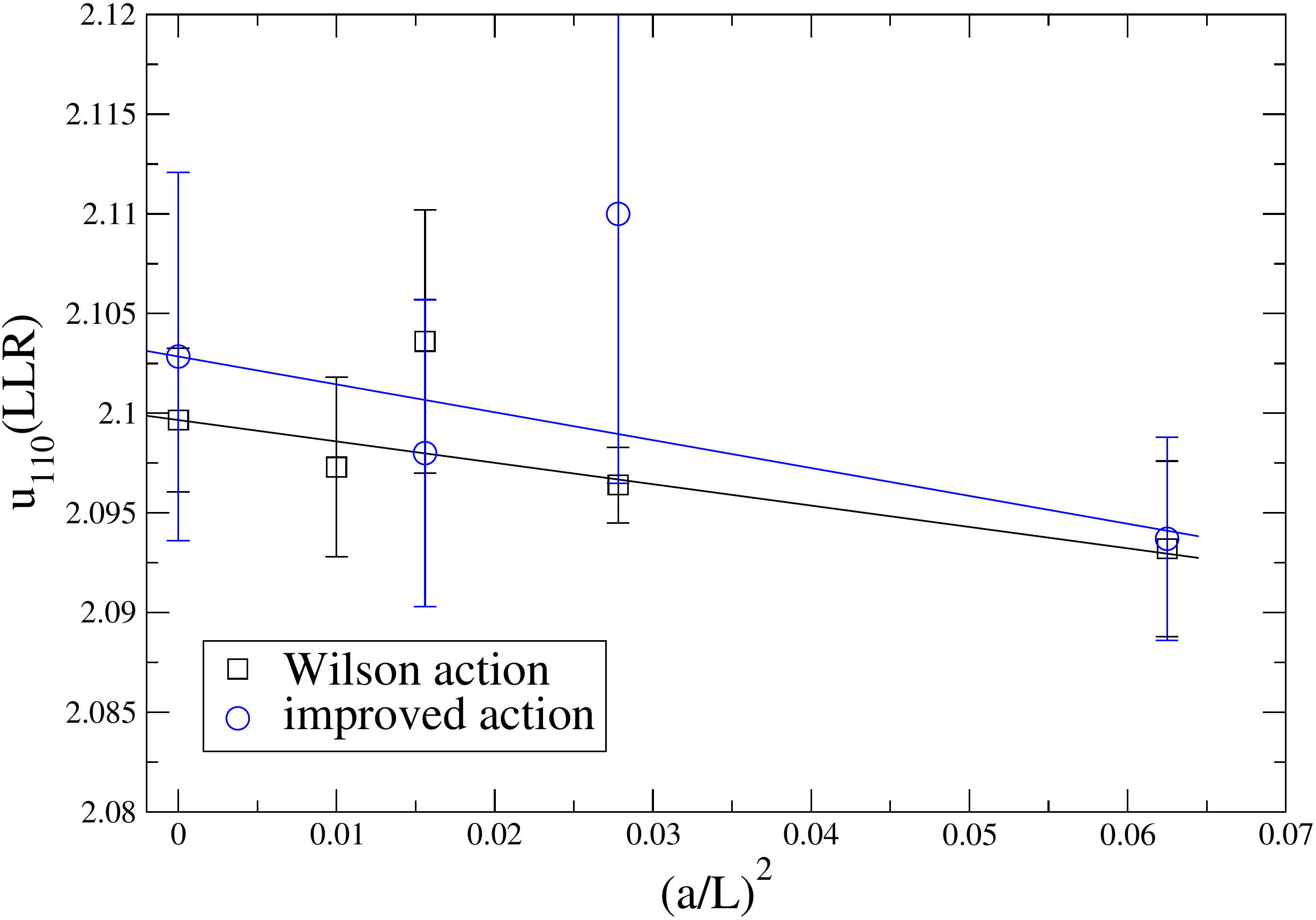}
  \caption{The $u_{110}$ torelon mass for SU(2) measured on spatial volumes
    (left) $(L,L, 3L/2)$ and (right) $(L,L, 2L)$.}
  \label{SU2_u110_Lxx}
\end{figure}

We have measured the torelon masses on asymmetric spatial lattices both for
the improved action and the standard Wilson action. Note that this is a
completely independent set of lattice simulations from the ones which were
used to tune the action parameters. We have considered shapes of type
$(L,L,3L/2), (L,3L/2,3L/2), (L,L,2L)$ and $(L,2L,2L)$, with the shorthand
notation $(LLr), (Lrr), (LLR)$ and $(LRR)$ in the plots and text.  The
corresponding results are shown in Figs.~\ref{SU2_u100_LLx}-\ref{SU2_u110_Lxx}
for torelons which wrap around either one or two of the short spatial
directions.  One could also try to detect lattice artifacts in the heavier
states such as the $u_{111}$ torelon in symmetric and asymmetric spatial
volumes, but we found these masses could not be extracted with sufficient
accuracy to be useful for comparison of cut-off dependence.

Let us first examine the $u_{100}$ states. As discussed earlier, the torelon
wrapping around the shortest distance gets lighter as the transverse spatial
directions increase in size. For example the $u_{100}(LRR)$ state is 2.5 times
lighter than $u_{100}(L) = 1.375$. This makes their determination from the
exponential decay of Polyakov loop correlators somewhat easier, as the signal
persists for larger time separation. We see that both for the improved and the
Wilson actions the lattice artifacts again appear to be ${\cal O}(a^2)$. Once
extrapolated to the continuum, there is very good agreement between the two
sets of simulations, except for some possible tension for $u_{100}(LRR)$. Note
that there is no tuning done at this stage: the bare couplings $\beta$ and
$\gamma$ have been fixed by requiring $u_{100}(L) = u^\star = 1.375$. The
improved action has consistently smaller lattice artifacts than the Wilson
action, and on the coarsest lattice $L/a = 4$, the cut-off dependence in
e.g.~the $u_{100}(LRR)$ is reduced from $\sim 25$\% with the Wilson action to
$\sim 4$\% for the improved action.

We next discuss the $u_{110}$ states. An interesting empirical observation is
that, while the $u_{100}$ masses approach the continuum limit from above, the
$u_{110}$ masses approach it from below. Unfortunately, the statistical precision is
not good enough to make a strong statement about reduced artifacts with the
new action at finite lattice spacing. It is important however that we find
consistency in the continuum results between the Wilson and improved action
simulations, which are both determined with better than 1\% precision.

Given that our tuning of the action parameters used $u_{100}(LLL)$ and
$u_{110}(LLL)$, both measured on cubic $L^3$ spatial volumes, one of the
limitations was the reduced statistical accuracy of the heavier $u_{110}$
state. An alternative strategy would be to fix the optimal couplings by
choosing the pair $u_{100}(LLL)$ and $u_{100}(Lrr)$, the second state having a
lighter mass and hence being easier to measure. However there is the drawback
that one needs two full sets of simulations on both symmetric and asymmetric
spatial lattices to complete the tuning.\footnote{
Another drawback is connected to our choice of $L$ being close to $1/T_c$.
In this case increasing the transversal size one gets closer to the critical
situation and the fluctuation of the Polyakov loops get larger.}
We pursued this approach and found the results were similar: 
with the new lattice action we could not completely eliminate the cut-off
dependence on the coarsest $L/a=4$ lattice. Hence we do not show these 
results here.

At this point it is interesting to note that the SU(2) case is special.
Considering a large spatial volume $L^3$, the flux tubes going along the
diagonal directions, say along $(1,1,0)$ and $(1,-1,0)$ are obviously two
different states. However, they belong to the same sector $\mathbf{k}=(1,1,0)$
in the SU(2) case.  
As a consequence, these states are mixed, and the eigenstates of the transfer
matrix are even/odd w.r.t.~$90^\circ$ rotation in the 1-2 plane.
The operators producing these mass eigenstates from the vacuum sector can be
constructed by
\begin{equation}
  \phi_{110}^{(\pm)}(t) = \Tr\left( \Phi_1(t) \Phi_2 (t)
    \pm \Phi_1(t) \Phi_2^\dagger(t) \right).
\end{equation}
One expects that the energy difference between the odd and even lowest states
is relatively large for small $L$ (where the width of the flux spreads over
the available volume, and there is a large overlap between fluxes going along
the two diagonals) and tiny for large $L$.  Note that the operator given in
\eqref{phik} is an even one. Similar eigenstates also appear in the 111 sector.
Although for some cases we measured the odd torelon masses like 
$m_{110}^{(-)}(L)$, we did not use them in this work.
 
\subsection{Static Potential}
\label{sec:potential}

In this section we present the scaling behavior of quantities related to the
static potential. Usually the force between quarks is used as a practical and
simple way to fix the scale, relating the bare coupling and the lattice
spacing in physical units~\cite{Sommer:1993ce}. Here we consider the reverse
approach. We investigate the approach to the continuum limit of the force
between static quarks after fixing the scale with the torelon mass.

We have performed numerical simulations of SU(2) Yang-Mills theory comparing
the standard Wilson action and the improved one. We have considered the bare
couplings tuned with the torelon mass $m_{100}(L)$ for $L=4$, $6$ and $8$
and listed in table~\ref{SU2_par}. The actual sizes of the simulated lattices
were $24^4$, $36^4$ and $36^3\times48$, respectively, for the three above
parameter sets. The static potential has been measured on axis from the
two-point correlation function of Polyakov loops at distance $r$
\begin{equation}
  V(r)=-\frac{1}{L_0} \mbox{log} \langle \Phi(0) \Phi(r) \rangle
\end{equation}
where $L_0$ is the lattice size along the temporal direction. The Monte Carlo
simulations have been carried out using the multi-level
algorithm~\cite{Luscher:2001up}. The force $F$ is obtained from the static
potential by
\begin{equation}
  F(r') = V(r) - V(r-1)\,,
\end{equation}
where $r'$ is a properly chosen point between $r$ and $r-1$.  The simplest
case is the midpoint $r'=r-1/2$. The cut-off effects can be somewhat
reduced by choosing $r'= r_I(r)$, the tree-level improved midpoint distance
between $r$ and $(r-1)$~\cite{Sommer:1993ce}.  We present here the naive
choice $r'=r-1/2$, but the qualitative behavior of the cut-off effect is the
same with the other choice as well.

\begin{figure}[tbp]
  \centering
  \includegraphics[width=.45\textwidth]{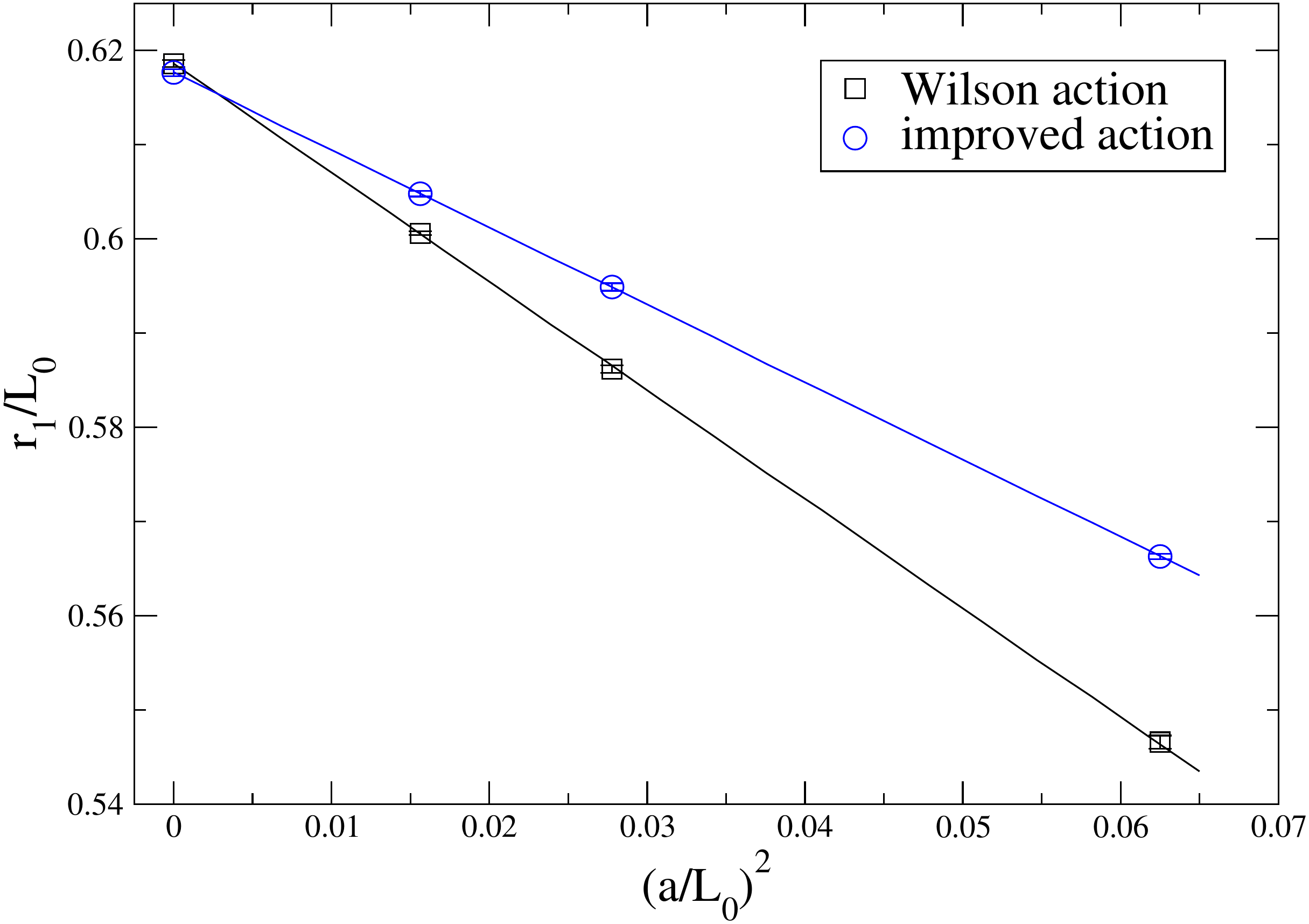}
  \hfill
  \includegraphics[width=.45\textwidth]{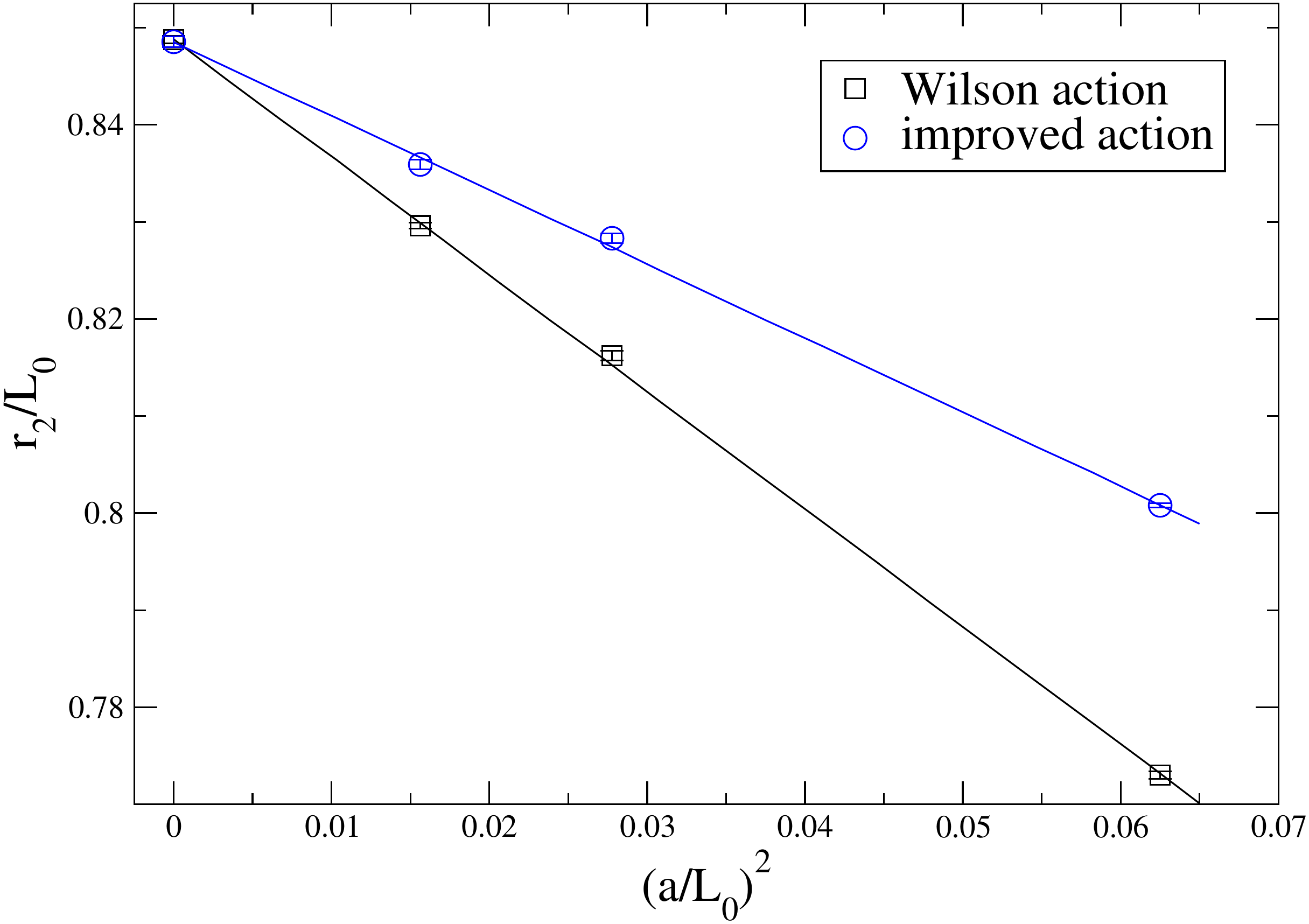}
  \caption{Approach to the continuum limit of $r_1/L_0$ (left) defined by
    $H(r_1) = 1$ and $r_2/L_0$ (right) defined by $H(r_2) = 1.65$ as a
    function of $(a/L_0)^2$.  $L_0$ is defined by $u_{100}(L_0)=1.375$\,.}
  \label{scalingH}
\end{figure}

The force is a dimensionful physical observable and it is useful to
investigate its scaling behavior by considering the dimensionless quantity
\begin{equation}
  H(r) = F(r) r^2.
\end{equation}
We have studied the approach to the continuum limit of two scales, $r_1$ and
$r_2$ defined as the distances where $H(r)$ has the values 1.0 and 1.65,
respectively. Note that -- as we mentioned above -- one usually fixes the
scale by defining the Sommer parameter, $r_0=0.5\,\mathrm{fm}$, as the
distance where $H(r_0)= 1.65$.  In figure~\ref{scalingH} we show the scaling
behavior of $r_1/L_0$ and $r_2/L_0$ as a function of $(a/L_0)^2$ where the
scale $L_0$ is defined by the torelon mass through $L_0
m_{100}(L_0)=1.375$. With the improved action, the reduction of lattice
artifacts in the static force is less pronounced than in the torelon
masses. For both actions the artifacts appear to be linear in $a^2$ and there
is very good agreement between the continuum extrapolation results.

{It is important to note
that the lattice artifacts of the static potential can be only partially
attributed to the properties of the lattice action.  The choice of the
operator (say smeared Polyakov loops versus naive ones) also contributes, and
it is not easy to separate these effects.  
Even in continuum electrostatics, the force between two charges of 
cubic shape is not exactly proportional to $1/r^2$, it also depends 
on the relative orientation of the cubes.

\section{The SU(3) case}
\label{sec:SU3}

Due to its obvious relevance as part of QCD, and as a possible prelude for
future work, we also tuned and tested the improved action in SU(3) gauge
theory, and compared it to the standard Wilson action to see how much the
cut-off effects could be suppressed. Because much of the procedure is similar
to the SU(2) work described above, we avoid a detailed description of the
common aspects.

We use the same parametrization of the gauge action as in
eq.~\eqref{impaction} and the same value $q=10$ to curb large fluctuations of
the plaquette. For the tuning procedure, we again choose to keep the physical
volume fixed in units of the torelon mass, with the precise value being
$m_{100}(L) L \equiv u_{100}(L) = u_{100}^\star = 1.0$. Note that for the same
value of $L/a$ this corresponds to a somewhat finer lattice spacing than in
our SU(2) study. Starting on the coarsest lattice $L/a = 4$, we tested a range
of values of $\gamma$, for each one finding the tuned coupling $\beta(\gamma)$
where the above condition was satisfied. We simultaneously measured the
$u_{110}$ torelon mass at the same $\gamma$ and $\beta(\gamma)$ values. We
found that for $\gamma = 200$ the cut-off effects in the $u_{110}$ state were
largely removed at this coarse lattice spacing. This corresponds to an
effective cut on the plaquette values at $\delta = \gamma^{-1/q} =
0.59$. Going to larger values of $\gamma$ significantly reduces the efficiency
of the Monte Carlo simulations, with decreasing improvement in reducing
lattice artifacts. For this reason, we used the fixed value $\gamma = 200$ for
the remainder of the SU(3) study.

\begin{table}[htb]
  \centering
  \begin{tabular}{rlcllll}
    $L/a$ & \phantom{9999}$\beta$ & $\gamma$ 
    & \phantom{9} $u_{100}(L)$ & \phantom{9} $u_{110}(L)$ &
    \phantom{9} $u_{110}^\star(L)$ & \phantom{9}$a$[fm] \\ 
    \hline
    4 &  \m5.7299  &  0. & 1.0000(11) & 2.0990(32) & 2.0990(37) & 0.159\\
    6 &  \m5.9806  &  0. & 1.0000(9)  & 2.1797(28) & 2.1797(33) & 0.096\\
    8 &  \m6.1749  &  0. & 1.0000(25) & 2.2050(53) & 2.2050(74) & 0.070\\
    10 &  \m6.3364  &  0. & 1.0000(36) & 2.2066(100)& 2.2066(126) & 0.056\\
    12 &  \m6.4741  &  0. & 1.0000(35) & 2.2141(110)& 2.2141(130) & 0.047\\
    \hline
    4 &  $-2.4104$  & 200. & 1.0000(28) & 2.1811(74) & 2.1811(96) & \\
    6 &  \m0.2174   & 200. & 1.0000(16) & 2.1965(37) & 2.1965(51) & \\
    8 &  \m1.4882   & 200. & 1.0000(71) & 2.2057(70) & 2.2057(89) & \\
    \hline
  \end{tabular}
  \caption{Parameters of the SU(3) action and $u_{100}(L)$, $u_{110}(L)$ 
    on cubic lattices for the Wilson and the improved actions.
    The column $u_{110}^\star(L)$ includes in its error the propagated
    uncertainty in determining the location of
    $u_{100}(L)=1$. For reference we include for the Wilson action
    the value of the lattice spacing as set via the Sommer parameter
    $r_0=0.5$~fm \cite{Necco:2001xg}.} 
  \label{SU3_par}
\end{table}

We repeated the tuning exercise for the improved action at $L/a = 6$ and 8, as
well as finding the corresponding bare couplings for the Wilson action over a
range of lattice sizes from $L/a = 4$ to 12.  Our procedure to tune the action
parameters was technically slightly different for the SU(3) case. We measured
simultaneously the torelon masses at $\sim 40$ nearby $\beta$ values, and
fitted the $\beta$-dependence by a 5-th order polynomial.  Then from these
fits we determined the appropriate $\beta$ value from $u_{100}(L)=u^\star$ and
used this to determine $u_{110}(L)$.  The errors were determined by a
bootstrap procedure.  To present the data in a similar way as in
table~\ref{SU2_par}, in table~\ref{SU3_par} we give the value $u_{110}(L)$
corresponding to ``fixed $\beta$'' and $u^\star_{110}(L)$ corresponding to
``fixed $u_{100}(L)=u^\star$''.

\begin{figure}[tbp]
  \centering
  \includegraphics[width=.45\textwidth]{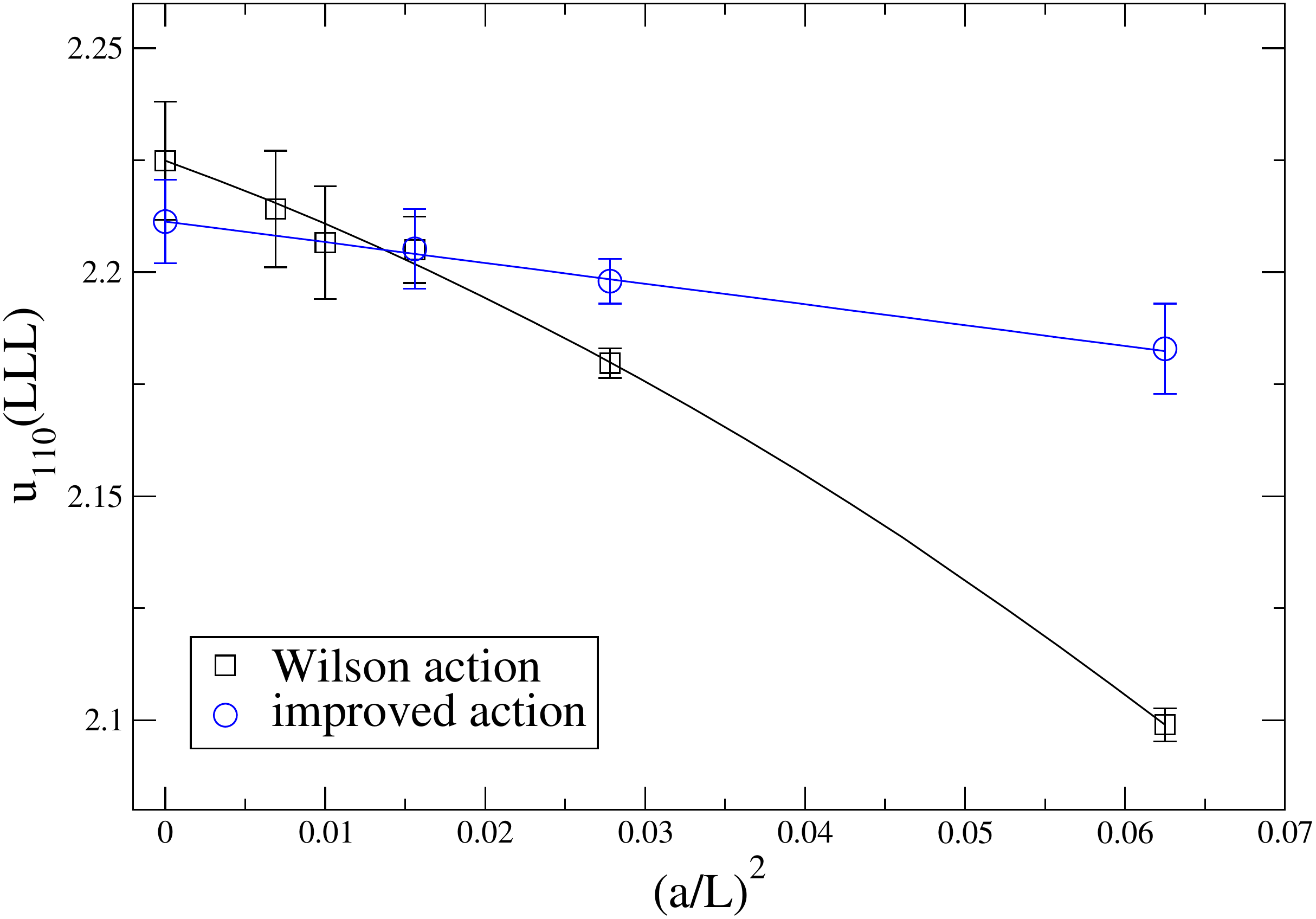}
  \caption{Cut-off effects of $u_{110}(L)$ on cubical lattices $L^3$ for the
    Wilson action and the improved action at $u_{100}(L)=1$ for SU(3).}
  \label{SU3_u110_LLL}
\end{figure}

We show the results for the $u_{110}(L)$ torelon mass for both improved and
Wilson actions in figure~\ref{SU3_u110_LLL}. At the coarsest lattice spacing,
the cut-off effects for the improved action are $\sim 1$\% compared to $\sim
5$\% for the Wilson action. Although both are small effects, at the level of
accuracy we could reach they are clearly observable. Similar to SU(2), by $L/a
= 6$ the improved action measurement lies almost on top of the continuum
extrapolated value. The overall smaller size of the cut-off effects compared
to the SU(2) case is possibly due to the choice of a smaller value of
$u_{100}^\star$. One contrast with respect to the SU(2) findings is that at
this level of accuracy, the continuum extrapolation must be quadratic in $a^2$
for the Wilson action if one wishes to include the $L/a = 4$ data point. For
the improved action, an extrapolation linear in $a^2$ describes the data
perfectly well.

\subsection{Scaling Tests}
\label{sec:SU3_tests}

As before for SU(2), once we complete the tuning procedure for $\gamma$ and
$\beta(\gamma)$ on symmetric spatial volumes, we next move to asymmetric
spatial volumes to examine what improvement the new action brings. The torelon
mass results are shown in Figs.~\ref{SU3_u100_LLx}-\ref{SU3_u110_LLx}. The
first question is whether or not there is good consistency between the
continuum extrapolated results using the two lattice actions, to which the
answer is yes. For the $u_{100}$ states the cut-off effects are
largely suppressed on the coarsest $L/a = 4$ lattice with the new action. For
the Wilson action on $L \times L \times 2L$ and $L \times 2L \times 2L$
volumes, again the continuum extrapolation of the $u_{100}$ mass is quadratic
in $a^2$ for the Wilson action, whereas the improved action extrapolation 
appears to be linear in $a^2$. 
The relative cut-off effect for the Wilson action increases as we go
towards the largest spatial volume $L \times 2L \times 2L$, at which point the
torelon is very light with $u_{100}(LRR) \approx 0.035$, given that the same
state on a symmetric volume corresponds to $u_{100} = 1$. The $u_{110}$ masses
are determined with 1\% accuracy or better, but as for SU(2), there is no
clear reduction of lattice artifacts with the newly proposed action.

\begin{figure}[tbp]
  \centering
  \includegraphics[width=.45\textwidth]{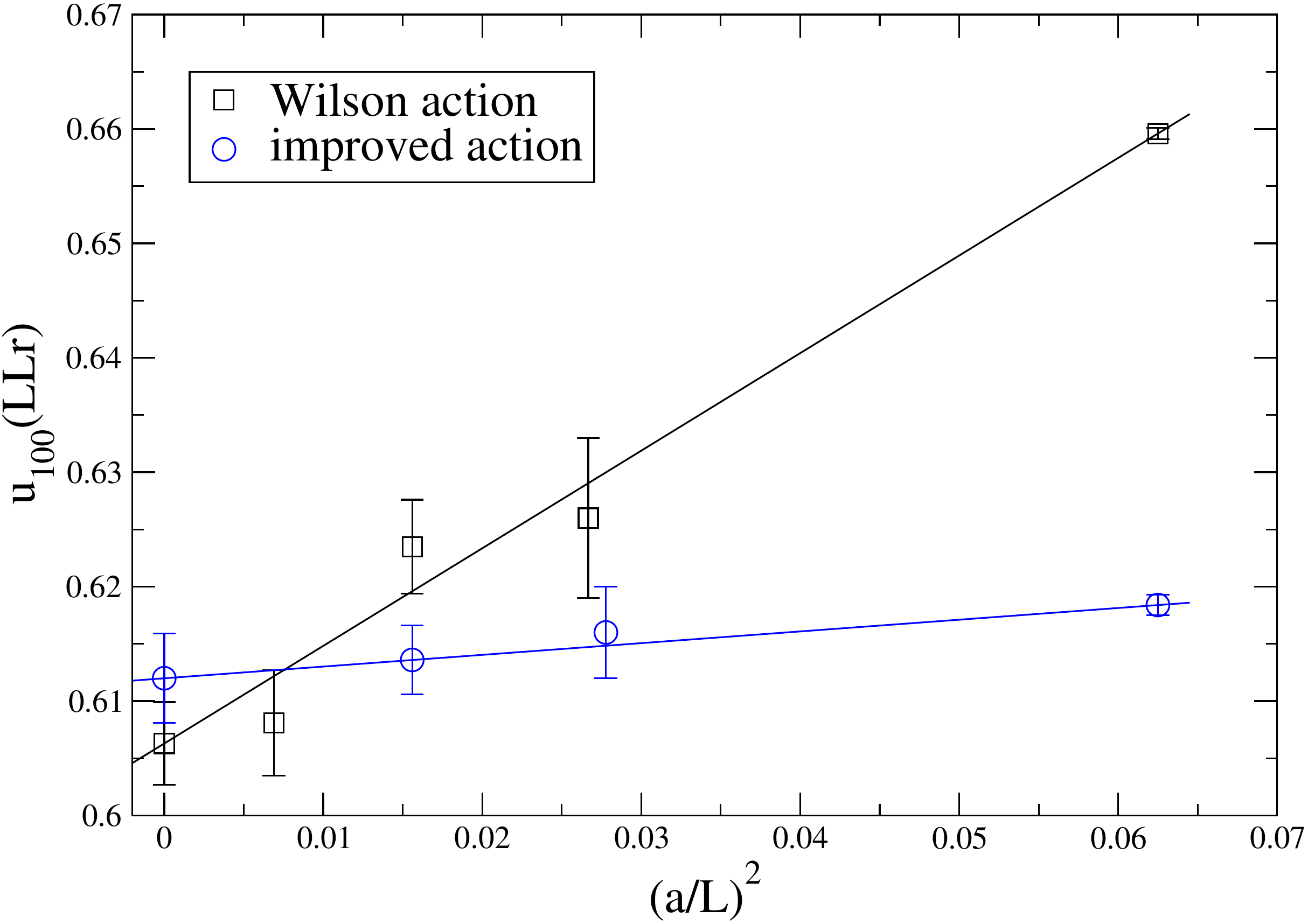}
  \hfill
  \includegraphics[width=.45\textwidth]{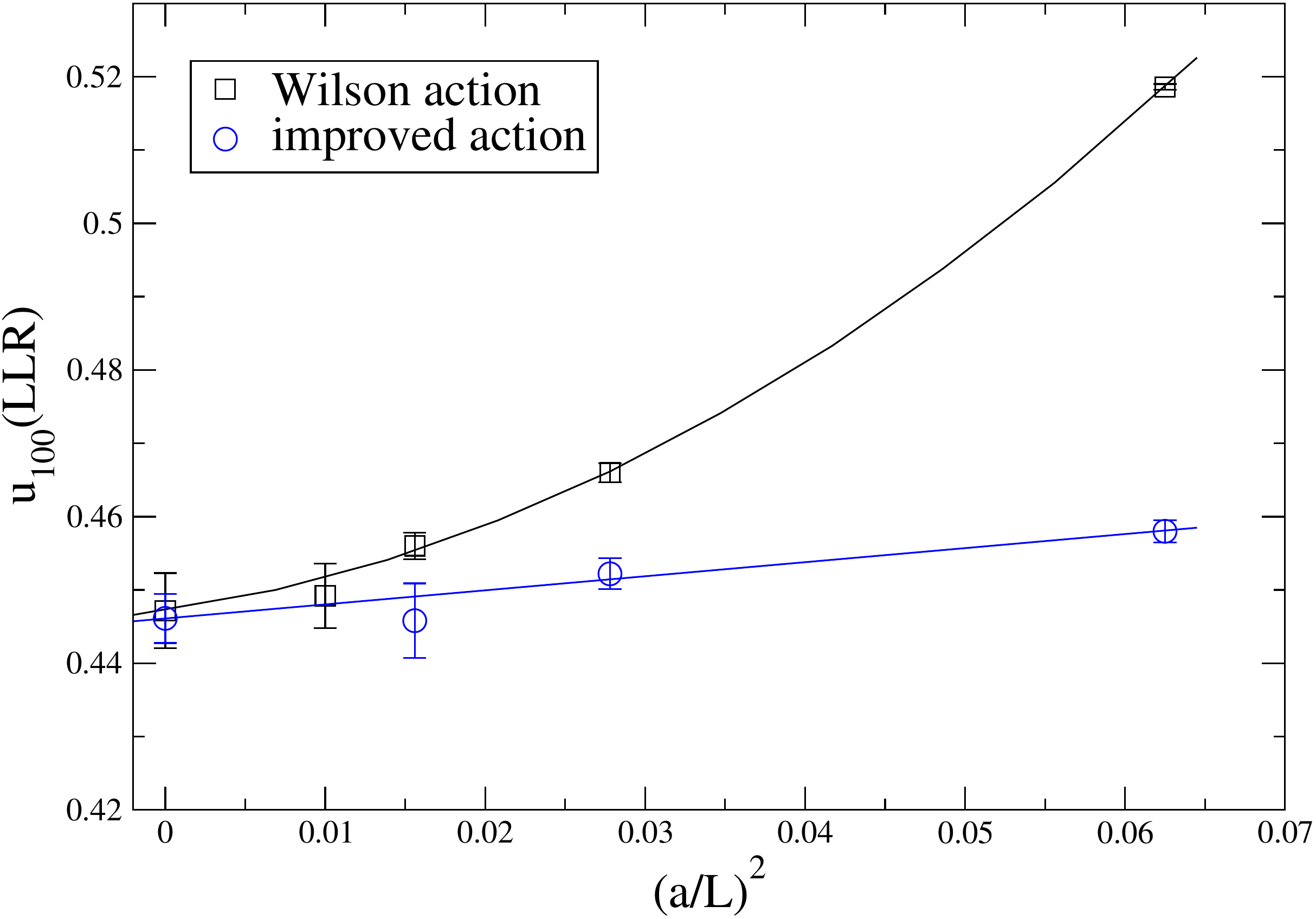}
  \caption{The $u_{100}$ torelon mass for SU(3) measured on spatial volumes
    (left) $(L,L, 3L/2)$ and (right) $(L,L, 2L)$.}
  \label{SU3_u100_LLx}
\end{figure}

\begin{figure}[tbp]
  \centering
  \includegraphics[width=.45\textwidth]{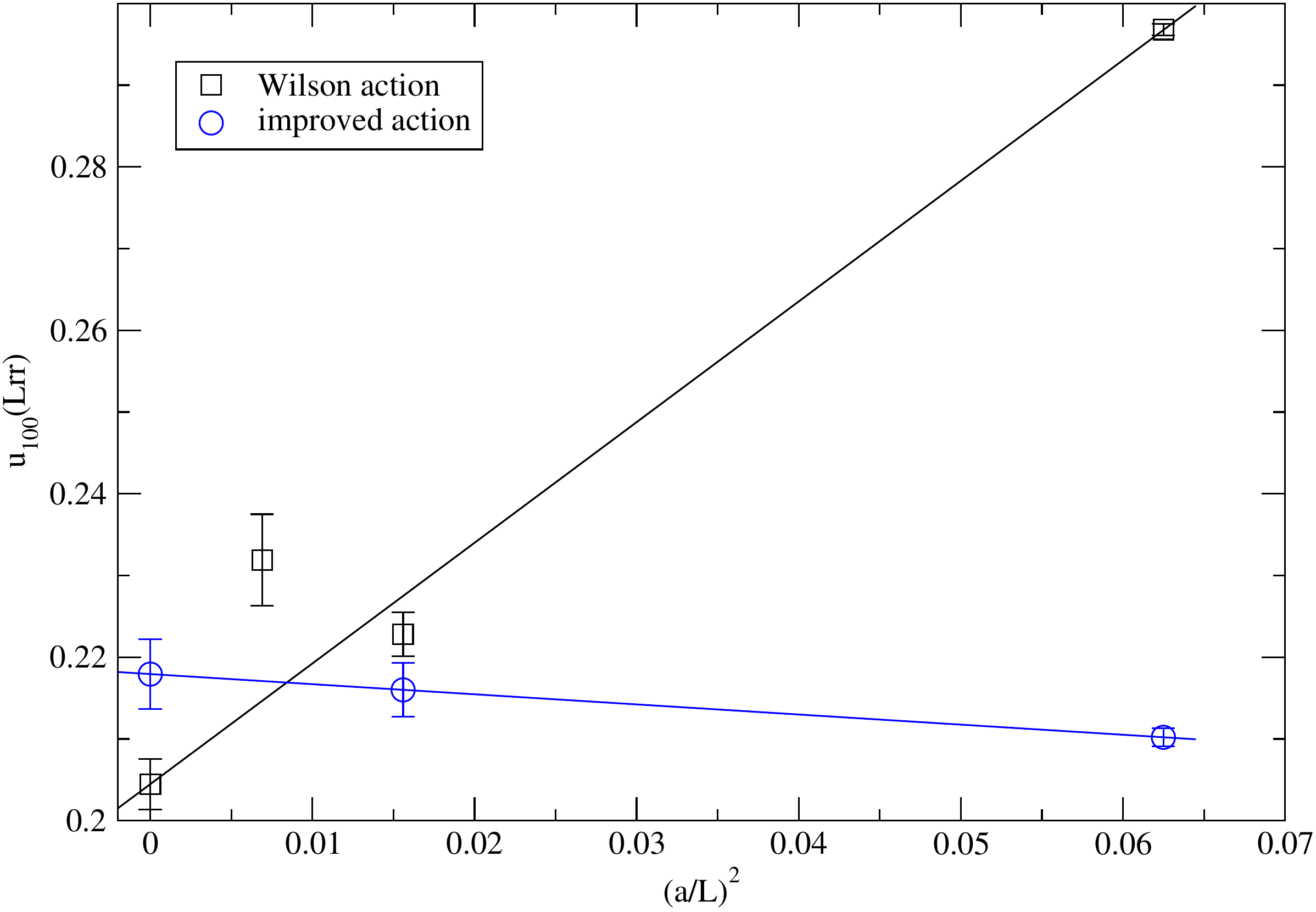}
  \hfill
  \includegraphics[width=.45\textwidth]{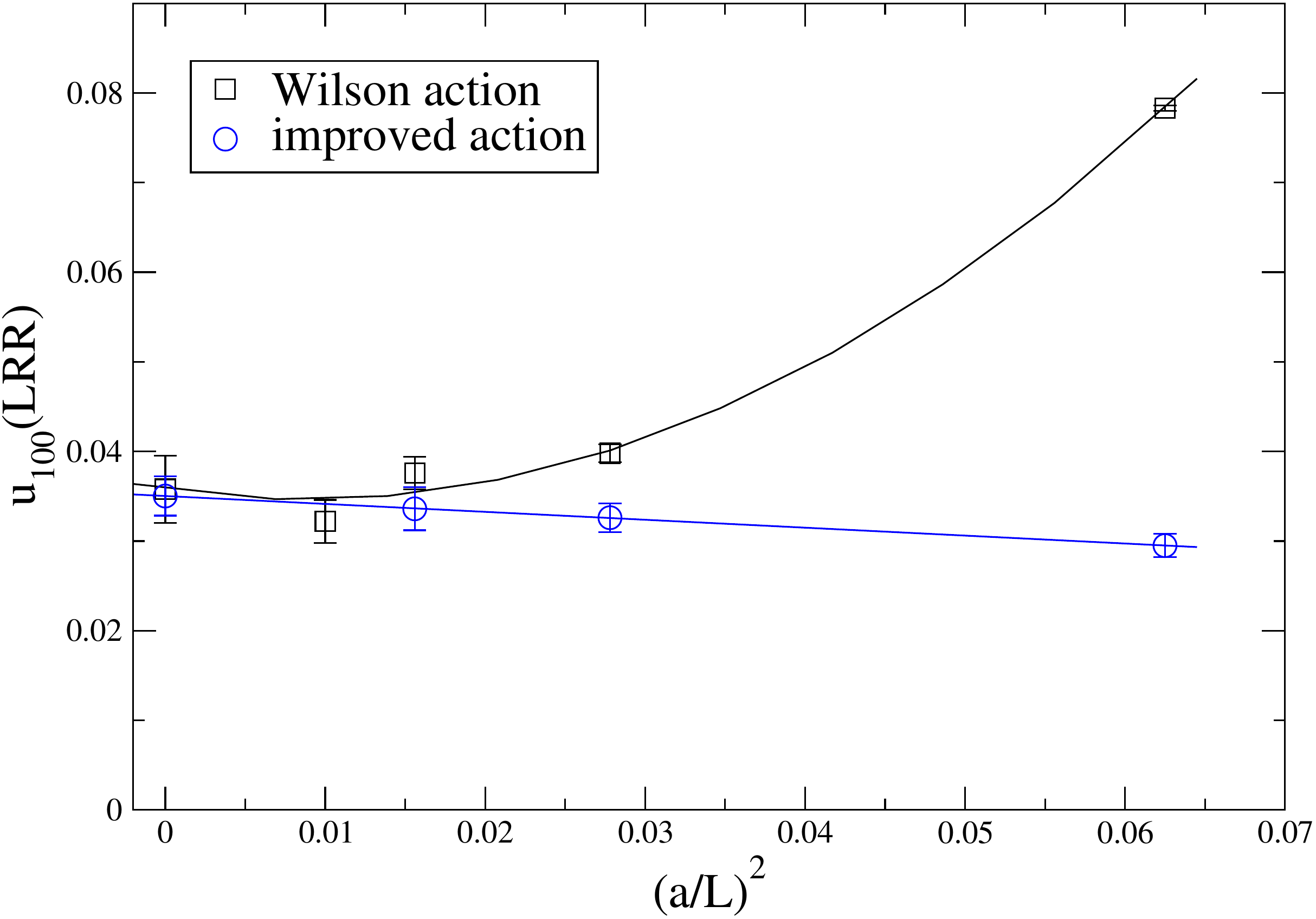}
  \caption{The $u_{100}$ torelon mass for SU(3) measured on spatial volumes
    (left) $(L,3L/2, 3L/2)$ and (right) $(L,2L, 2L)$.}
  \label{SU3_u100_Lxx}
\end{figure}

\begin{figure}[tbp]
  \centering
  \includegraphics[width=.45\textwidth]{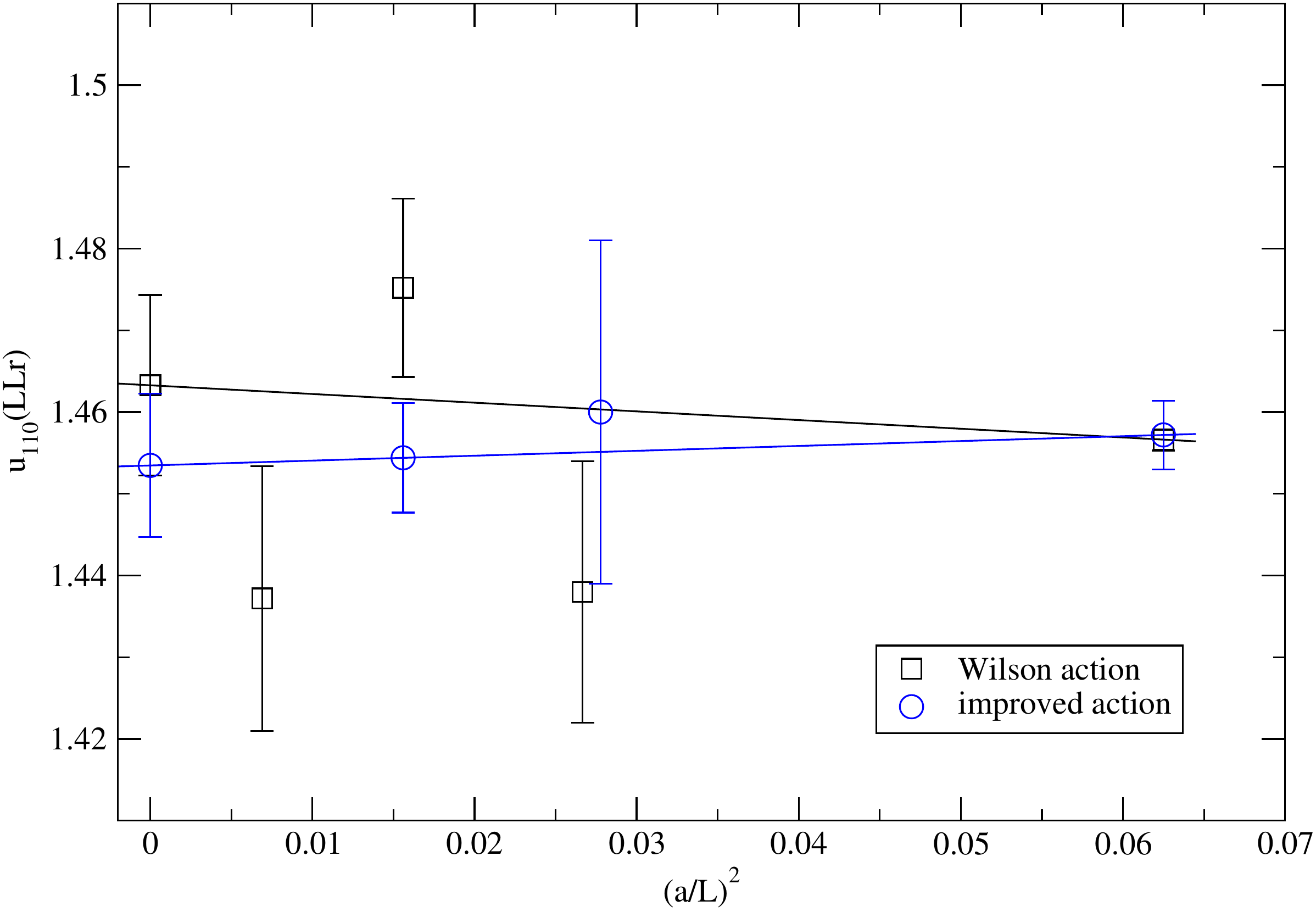}
  \hfill
  \includegraphics[width=.45\textwidth]{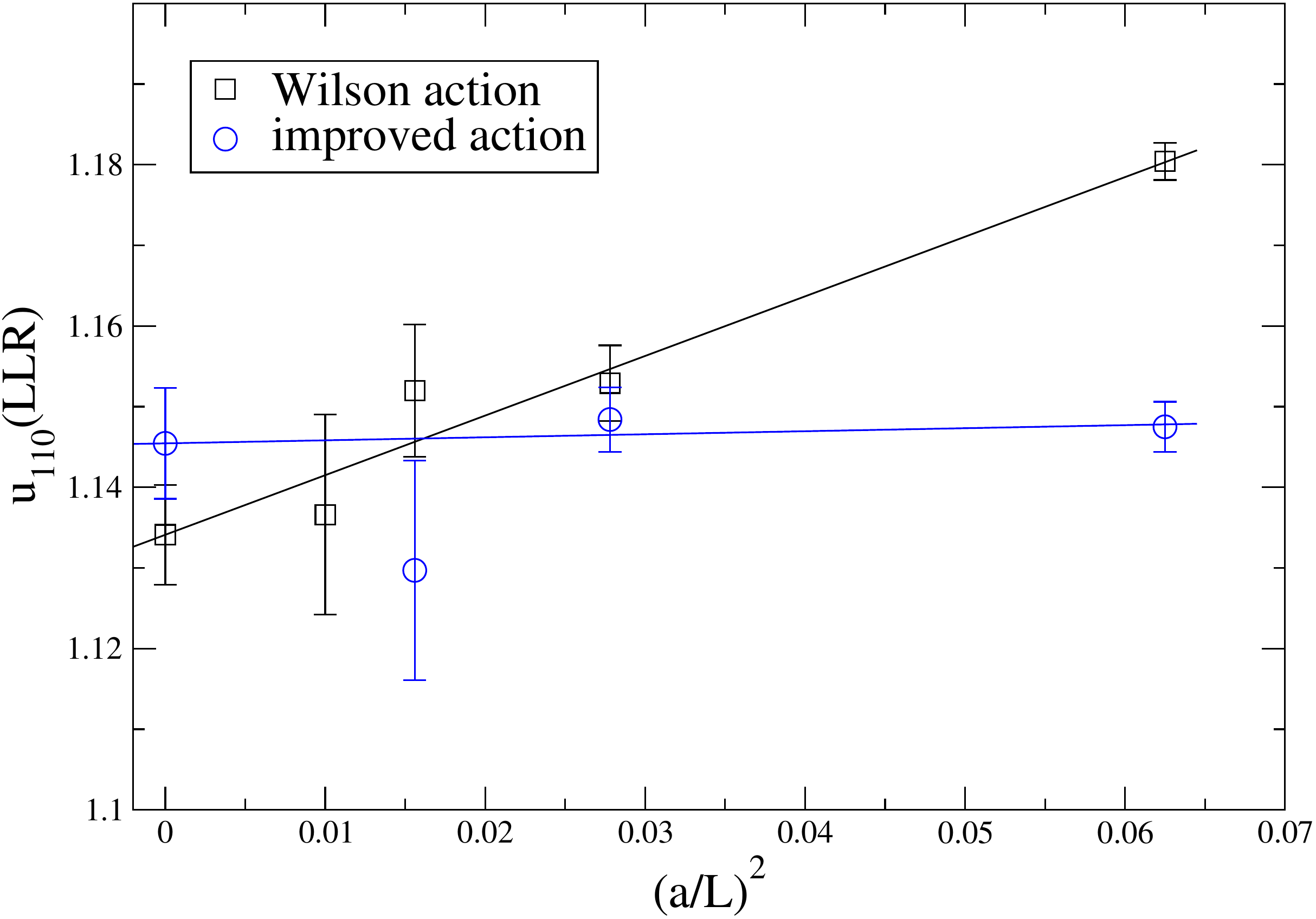}
  \caption{The $u_{110}$ torelon mass for SU(3) measured on spatial volumes
    (left) $(L,L, 3L/2)$ and (right) $(L,L, 2L)$.}
  \label{SU3_u110_LLx}
\end{figure}

\new{
To compare these lattice artifacts to those for the mixed fundamental-adjoint 
action we also measured the torelon masses for the action used in 
\cite{Hasenbusch:2004yq}.
Fixing the adjoint coupling to $\beta_a=-4.0$ as in \cite{Hasenbusch:2004yq}, 
with the fundamental coupling $\beta_f=9.398$ for $L/a=4$ we obtained
$u_{100}(L)=1.0036(8)$ and $u_{110}(L)=2.1502(27)$. This yields
the extrapolated value $u_{110}^\star(L)=2.1435(31)$, which is
halfway between values for the Wilson action and the newly proposed action 
(cf. table~\ref{SU3_par} and figure~\ref{SU3_u110_LLL}).
Hence the mixed fundamental-adjoint action brings some improvement, but not as much as the new action we present.
}

\subsection{Gradient flow observables}

The gradient flow is a recently developed method which smooths out lattice
fields in a controlled fashion and from which renormalized observables
can be measured with very high accuracy
\cite{Luscher:2010iy}. 
In the following we make use of the fact that the gauge field obtained at 
flow time $t/a^2 > 0$ is a smooth renormalized field \cite{Luscher:2011bx}. 
Hence, the expectation values of local gauge invariant expressions in 
this field are well-defined physical quantities that probe
the theory at length scales on the order of $\sqrt{t}$.  In particular, we
will consider observables related to the 
action density $E(t)$
at flow time $t$, and the set of its derivatives
\[
W^{(n)}(t) = (t \cdot \partial_t)^n \left(t^2 E(t)\right) \, .
\]
One possible discretization of the action density on the lattice
makes use of the sum of unoriented plaquettes with
a common lower-left corner \cite{Luscher:2010iy} and is denoted
by $E_\text{plaq}(t)$, but we
also used a more symmetric clover-type discretization
$E_\text{sym}(t)$. 

One way to exploit the gradient flow is to extract the lattice
spacing.  For example, one can define the lattice scale
$\sqrt{t_0}$ via the
requirement that
\begin{equation}
   W^{(0)}_0 \equiv \left. t^2 \langle E (t) \rangle \right|_{t_0} = c, 
  \label{eq:t2E}
\end{equation} 
where in the original investigation  \cite{Luscher:2010iy} the value $c = 0.3$
was chosen. The corresponding value of $t_0/a^2$ can be measured with very 
good precision and be used as a reference scale when taking the continuum 
limit $a\rightarrow 0$. 
Similarly one can define a scale $w_0$ set by the requirement that
\begin{equation}
  W^{(1)}_0 \equiv \langle W^{(1)}(t) \rangle _{t=w_0^2} = c', 
  \label{eq:Wn}
\end{equation}
where again there is freedom in the choice of the parameter $c'$, which in the
literature has first been tested for the choice $c' = 0.3$
\cite{Borsanyi:2012zs}. Note that for a given choice of $c$ and $c'$, the
dimensionless ratio $t_0/w_0^2$ is physical and has a well-defined value in the
continuum limit. Further derivatives of the action density renormalized at flow
time $t_0$ can be determined by evaluating
\begin{equation}
  W^{(n)}_0 \equiv \langle W^{(n)}(t_0,L) \rangle 
  \label{eq:Wnder}
\end{equation}
and the continuum limit of these quantities is taken by $\lim_{a\rightarrow 0}
W^{(n)}_0$.

Unlike the mass spectrum of the theory, gradient flow observables are not
spectral quantities, therefore their cut-off dependence follows from {\bf (a)}
the lattice action used in the Monte Carlo generation of the ensembles, {\bf
  (b)} the action used in implementing the flow,
and {\bf (c)} the lattice discretization of the action density. We generated
separate ensembles using both the standard Wilson and improved lattice gauge
actions over a range of lattice volumes
and lattice spacings. For both sets, we use the Wilson action in the flow 
and we consider both
$E_\text{plaq}$ and $E_\text{sym}$ to facilitate the continuum limit.
It turns out that the gradient flow observables in lattice units 
are essentially independent of the volume above 
$L/\sqrt{t_0} \gtrsim 8.0-8.5$.
As a consequence, we restricted our analysis of the results to simulations
for which $L/\sqrt{t_0} > 10.0$.}

\begin{figure}[thb]
  \centering
  \includegraphics[width=0.49\textwidth]{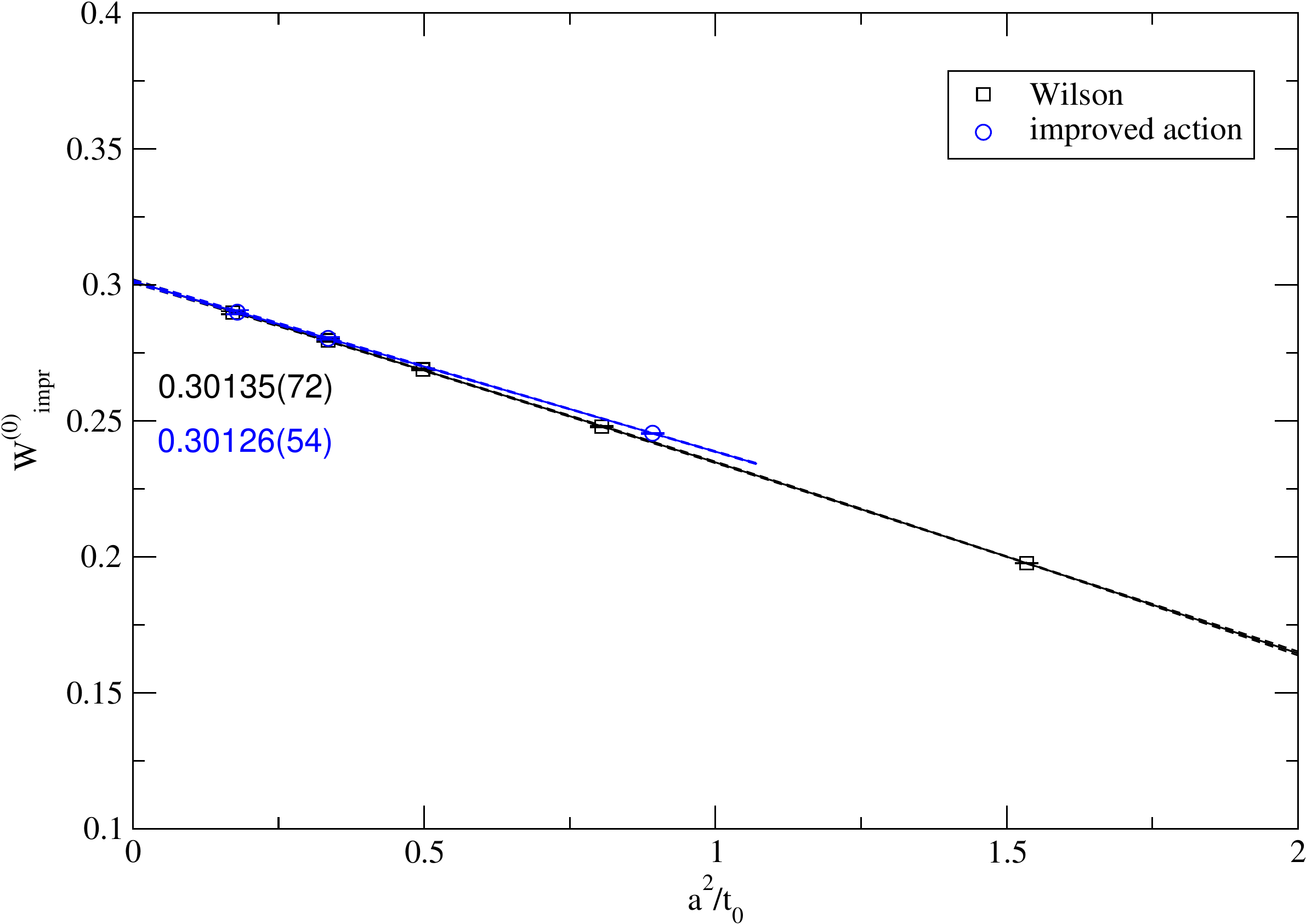}
  \includegraphics[width=0.49\textwidth]{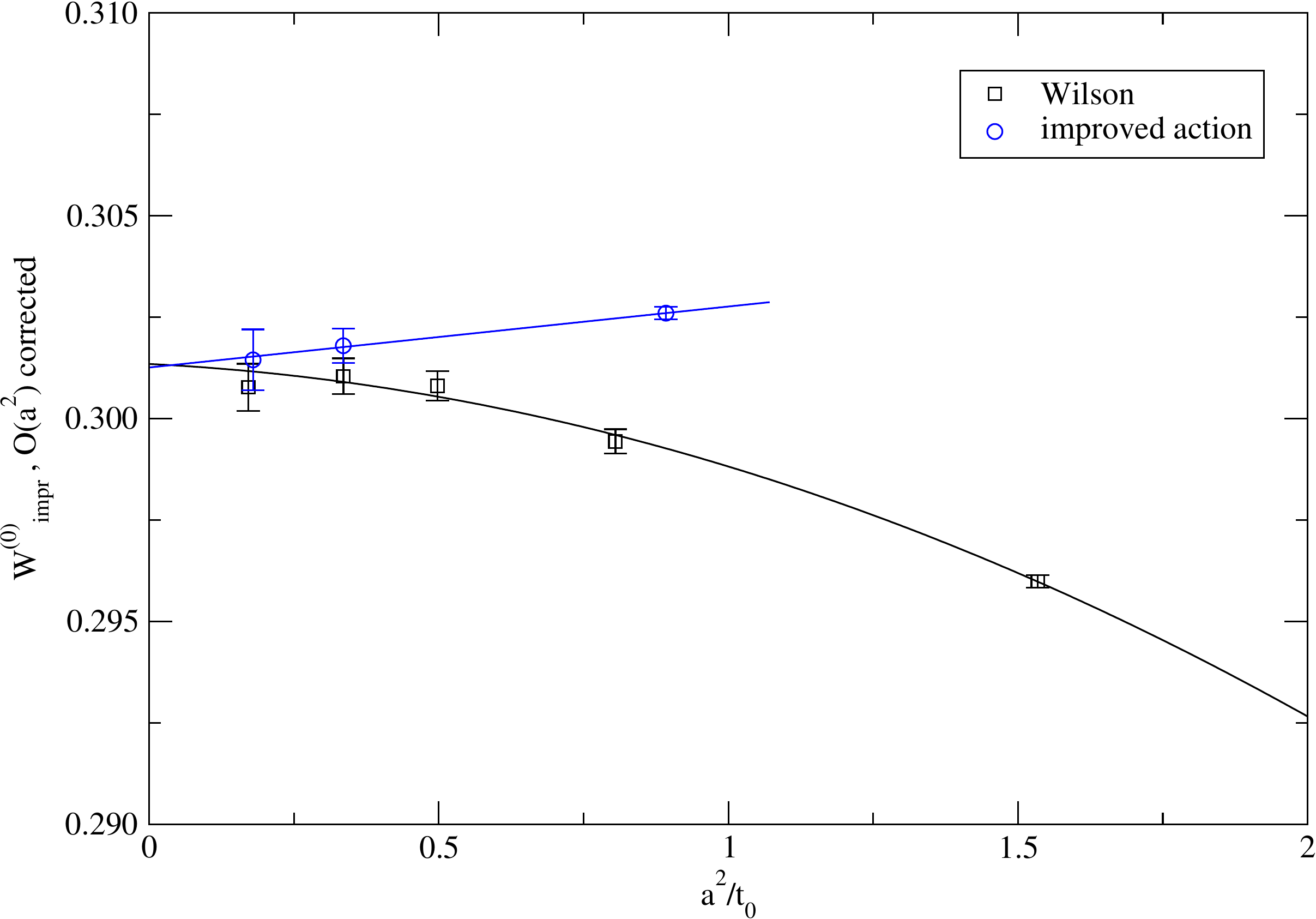}
  \caption{Continuum limit of $W^{(0)}(t_0)$ using the symmetrized definition
    for the action density while $t_0$ is determined from the plaquette
    definition. Original data (left) and ${\cal O}(a^2)$ corrected
    data (right).\label{fig:scaling_w0impr}}
\end{figure}

The quantification of the lattice artifacts due to the specific choice
of the gradient flow action is beyond the
scope of the present work, but the ones due to the
 discretization of the action density operator can be estimated by
comparing the observables evaluated with $E_\text{plaq}$ and $E_\text{sym}$.
In figure \ref{fig:scaling_w0impr} we show
the continuum limit of $W^{(0)}(t_0)$ using the symmetrized definition for the
action density while $t_0$ is determined from the plaquette definition. By
construction $W^{(0)}(t_0)$ takes the value $0.3$ in the continuum,
independent of the discretization employed for the action density. The
continuum limits displayed in figure \ref{fig:scaling_w0impr} show
that within two standard deviations this
is indeed the case for our simulation results, both for the Wilson and the
improved gauge action.
In order to enhance the differences between the two actions, we
removed the average ${\cal O}(a^2)$ correction which one can assume
stems from the discretization of the action density. 
We fit the results of the Wilson and improved action separately with 
an ansatz $c_0 + c_1a^2 + c_2a^4$, then average $c_1$ from both fits, 
and subtract $c_{1,ave}a^2$ from the original data. This is a convenient 
way to magnify the deviations. The result is displayed in the right plot 
of figure \ref{fig:scaling_w0impr} and illustrates that the continuum 
extrapolations can be achieved by employing an ${\cal O}(a^2)$ correction 
for the improved action, while ${\cal O}(a^2)$ and ${\cal O}(a^4)$ 
corrections are necessary for the Wilson action.
However, it is clear that the differences
between the results from the Wilson and the improved gauge action are
very small. We assign this to the fact that the lattice artifacts
introduced through the discretization of the operator and/or the
choice of the flow procedure are so large that they dominate the ones
due to the choice of the action for Monte Carlo simulation. 

\begin{figure}[thb]
  \centering
  \includegraphics[width=0.49\textwidth]{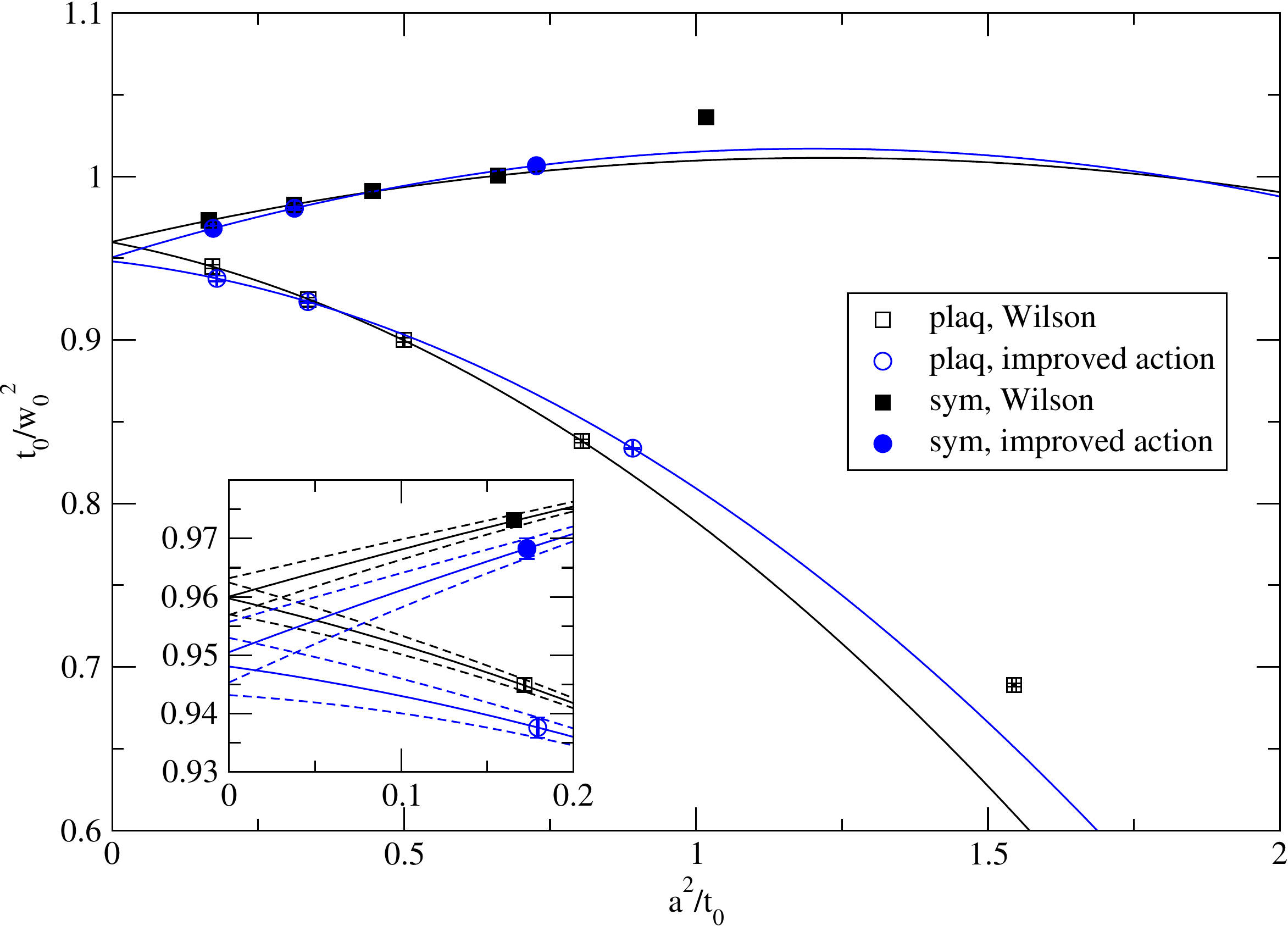}
  \includegraphics[width=0.49\textwidth]{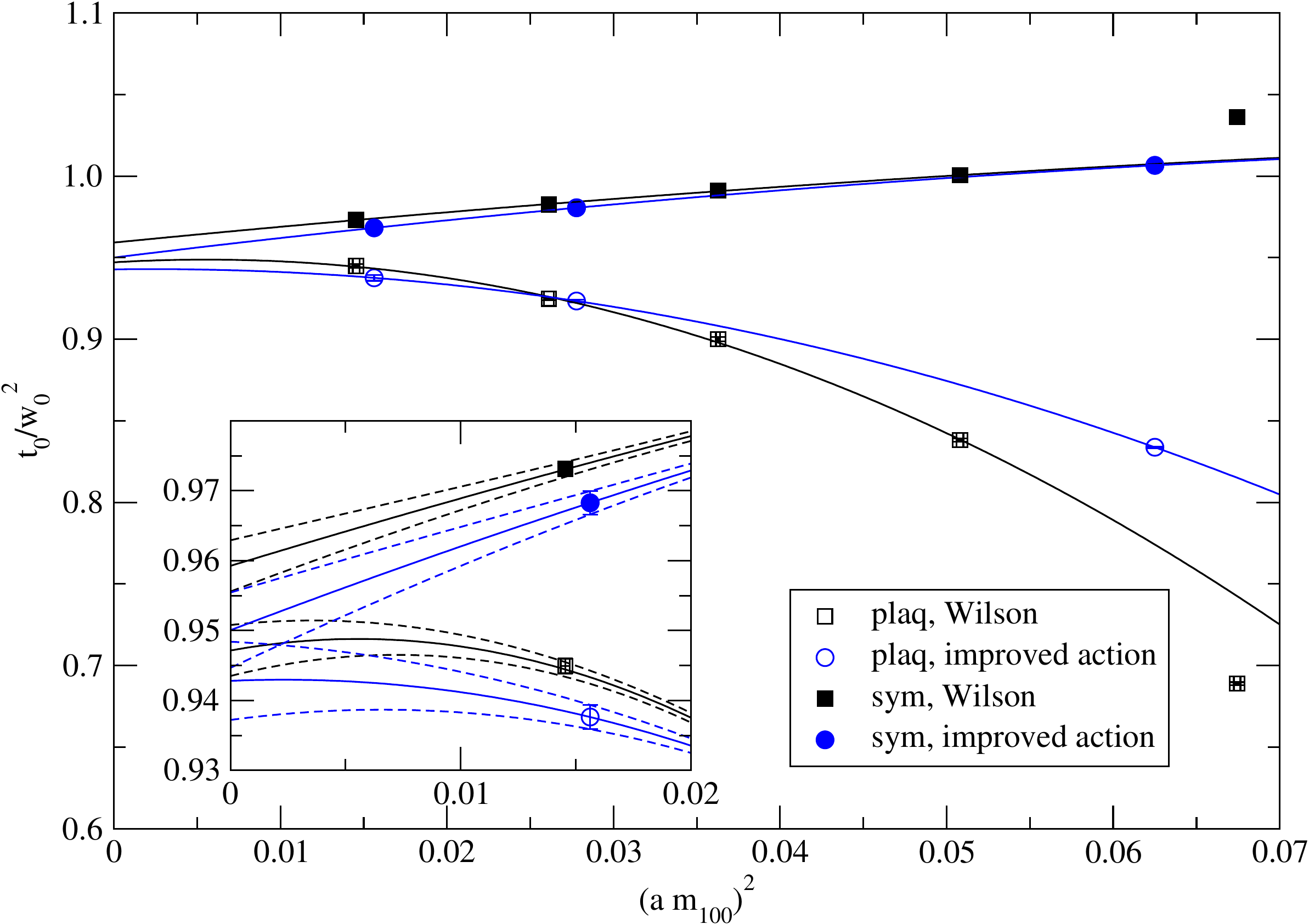}
  \caption{Continuum limit of the dimensionless ratio $t_0/w_0^2$ using the
    plaquette and the symmetrized definition for the action density in
    terms of $a^2/t_0$ (left
    plot) and in terms of $(a m_{100})^2$ (right plot).
    \label{fig:Rt0w02_vs_Ra2t0}}
\end{figure}

Yet another part of the
challenge in quantifying the magnitude of cut-off effects is the
ambiguity in the choice of lattice scale. To illustrate this, in
figure~\ref{fig:Rt0w02_vs_Ra2t0} we
consider the dimensionless ratio $t_0/w_0^2$ as a function of the lattice 
spacing expressed in units of $t_0$ (left plot) and in units of the torelon
mass $m_{100}$ (right plot) determined earlier in the spectroscopy study. 
On the coarsest lattices, corresponding to $am_{100} = 1/4$ for the new action, 
cut-off effects are now significantly smaller with the improved action. Still,
the dominant effect at finite lattice spacing is the choice of the action 
density operator. We tried to separate out the effect of the operator 
by considering a linear combination of $E_\text{plaq}$ and $E_\text{sym}$, 
so as to reduce the lattice artifacts stemming from the discretization 
of the action density operator. However, even in the improved combination 
the difference between the standard and new lattice actions is not dramatic.

\begin{figure}[thb]
  \centering
  \includegraphics[width=0.49\textwidth]{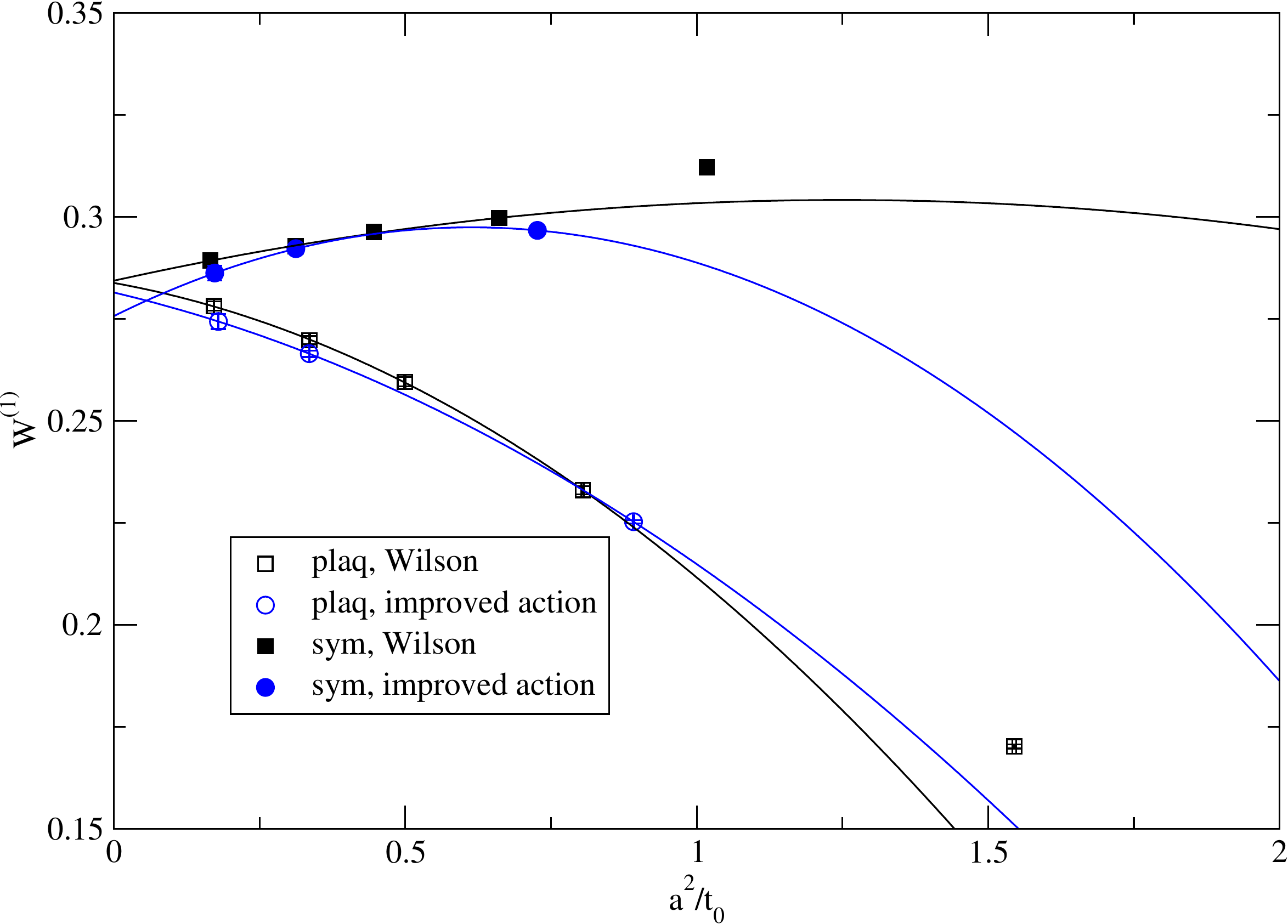}
  \includegraphics[width=0.49\textwidth]{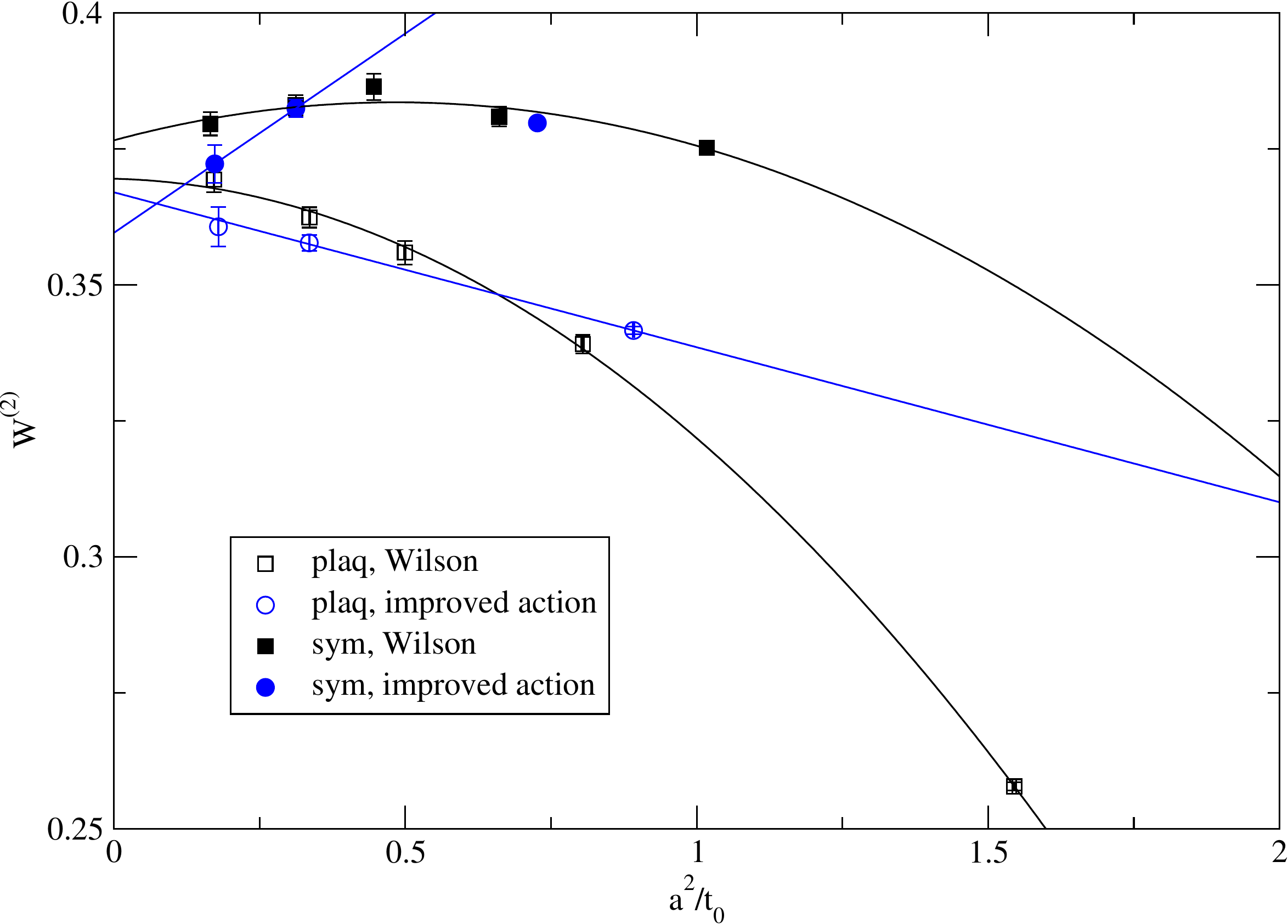}
  \caption{Continuum limit of the dimensionless derivatives $W^{(1)}(t_0)$ and
    $W^{(2)}(t_0)$ using the plaquette and the symmetrized definition for the
    action density.\label{fig:Wn_vs_Ra2t0}}
\end{figure}

Given the high precision of the gradient flow method, it is interesting to
investigate how many derivatives of the action density can be accurately
measured. In figure~\ref{fig:Wn_vs_Ra2t0} we show the continuum limit of the
first and second derivatives of the action density calculated from both the
plaquette and symmetrized definition, and for both gauge actions, with $t_0$
always being determined from the corresponding definition. We see that lattice
artifacts remain large and are again dominated by the choice of operator, not
the lattice action used to generate the ensemble. Most important, we see
consistent continuum results across the various discretizations, and cut-off
effects which are always well described by ${\cal O}(a^2)$ and ${\cal O}(a^4)$
corrections.

\subsection{Plaquette distribution}
\begin{figure}[thb]
  \centering
  \includegraphics[width=0.49\textwidth]{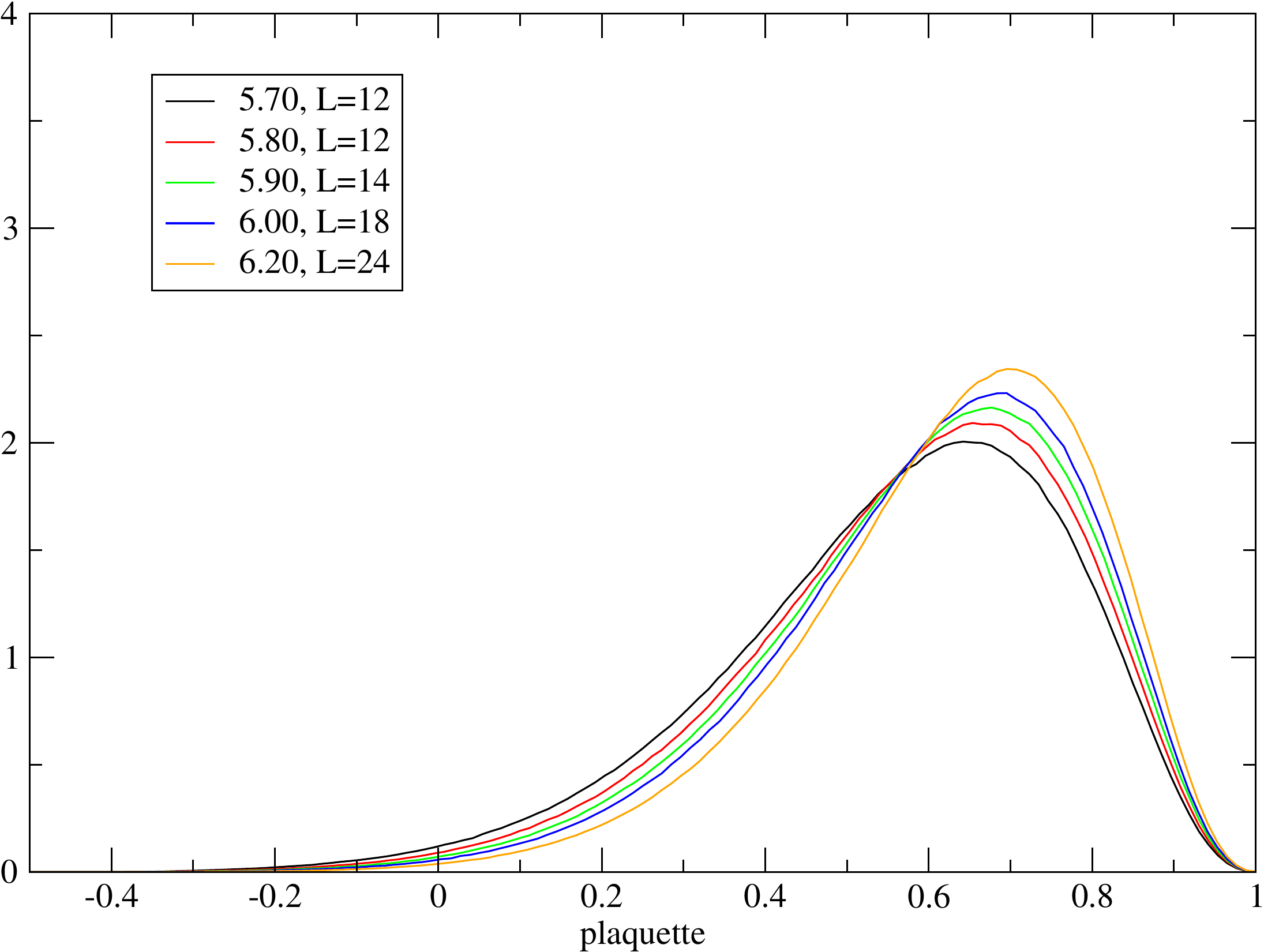}
  \includegraphics[width=0.49\textwidth]{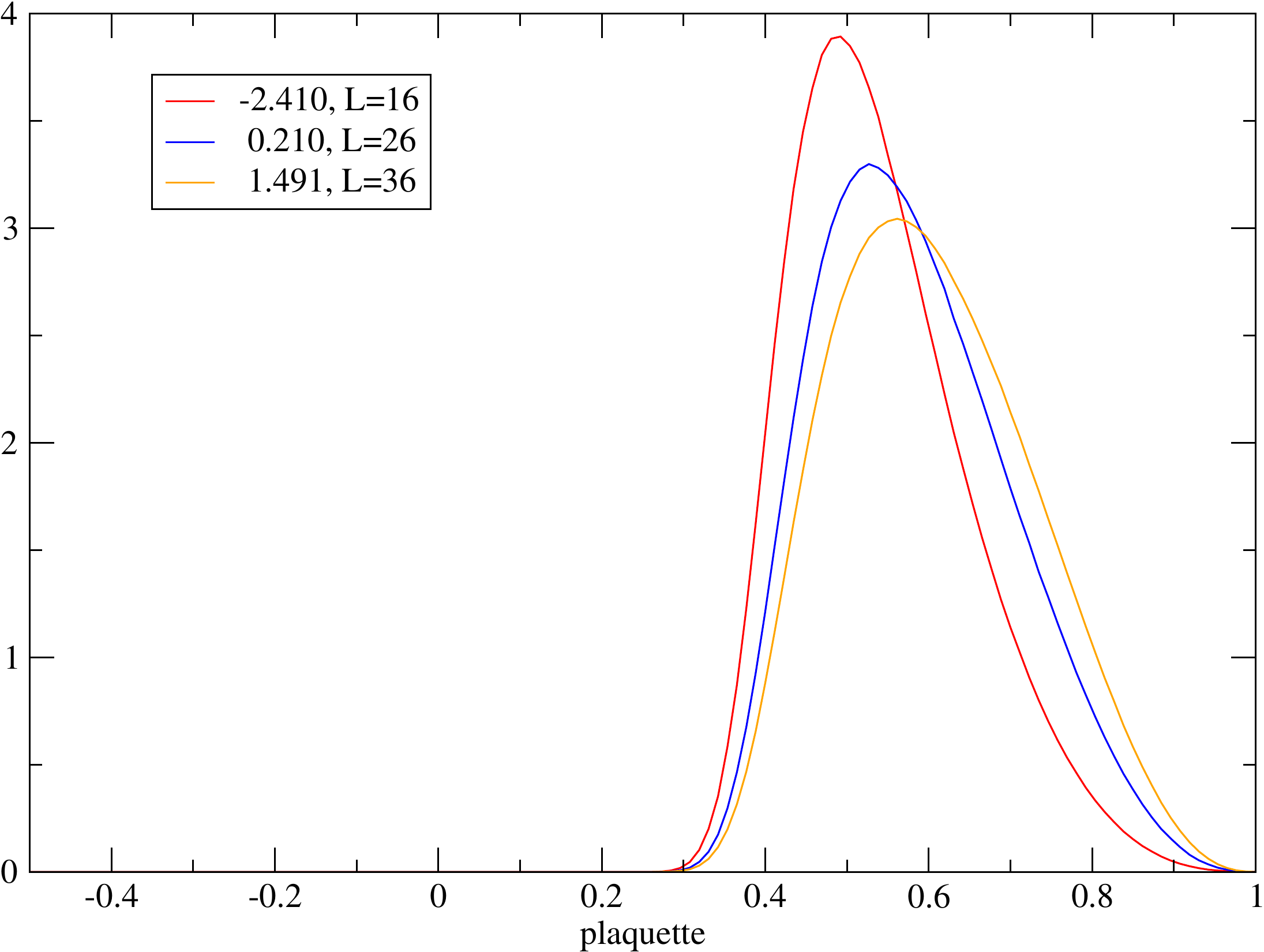}
\caption{Distribution of the plaquette values of the original gauge
  field configurations. The left panel corresponds to the Wilson ensembles
while the right panel shows the improved action ensembles.
\label{fig:plaquette_distr_orig}}
\end{figure}

\begin{figure}[thb]
  \centering
  \includegraphics[width=0.49\textwidth]{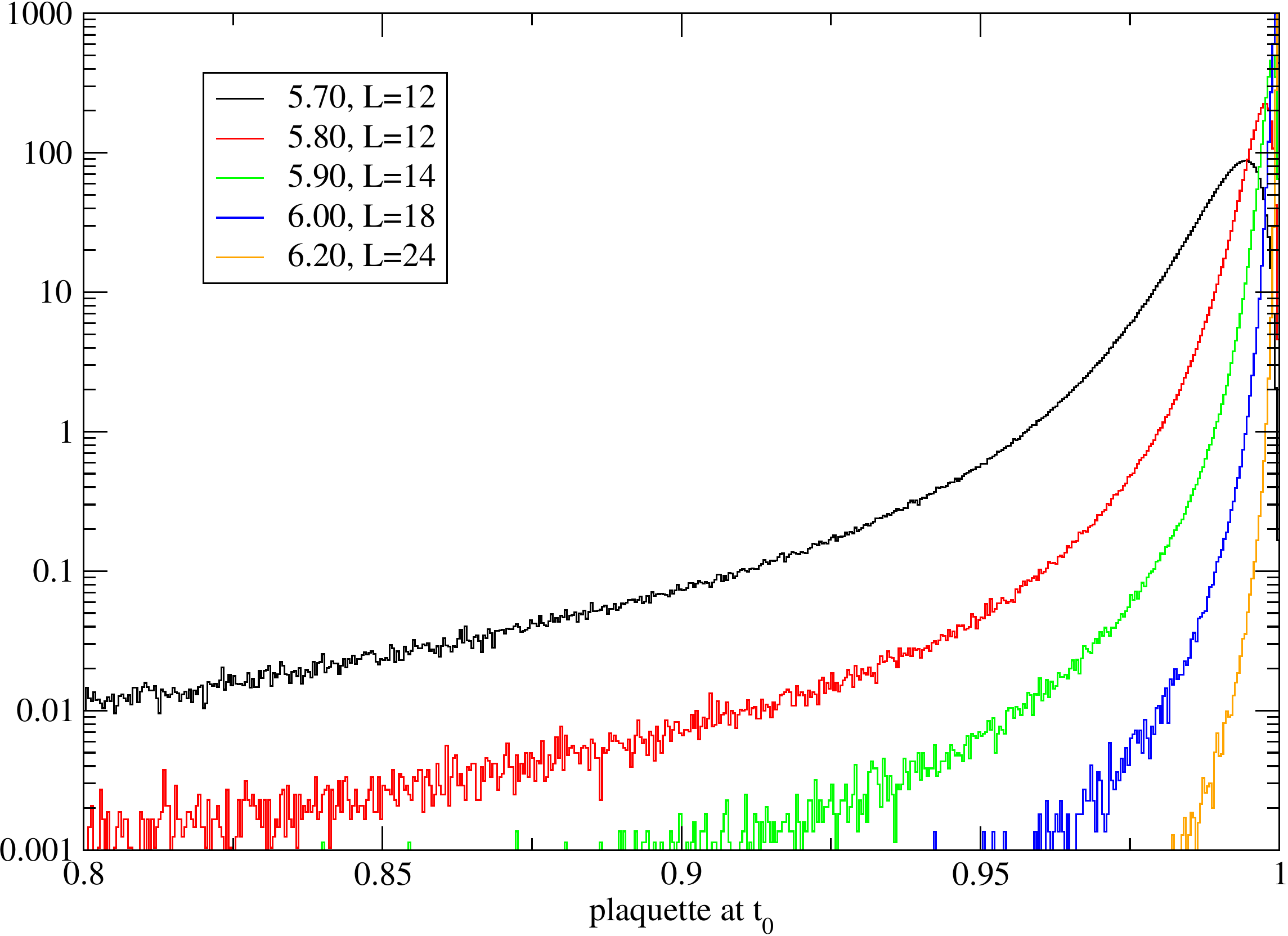}
  \includegraphics[width=0.49\textwidth]{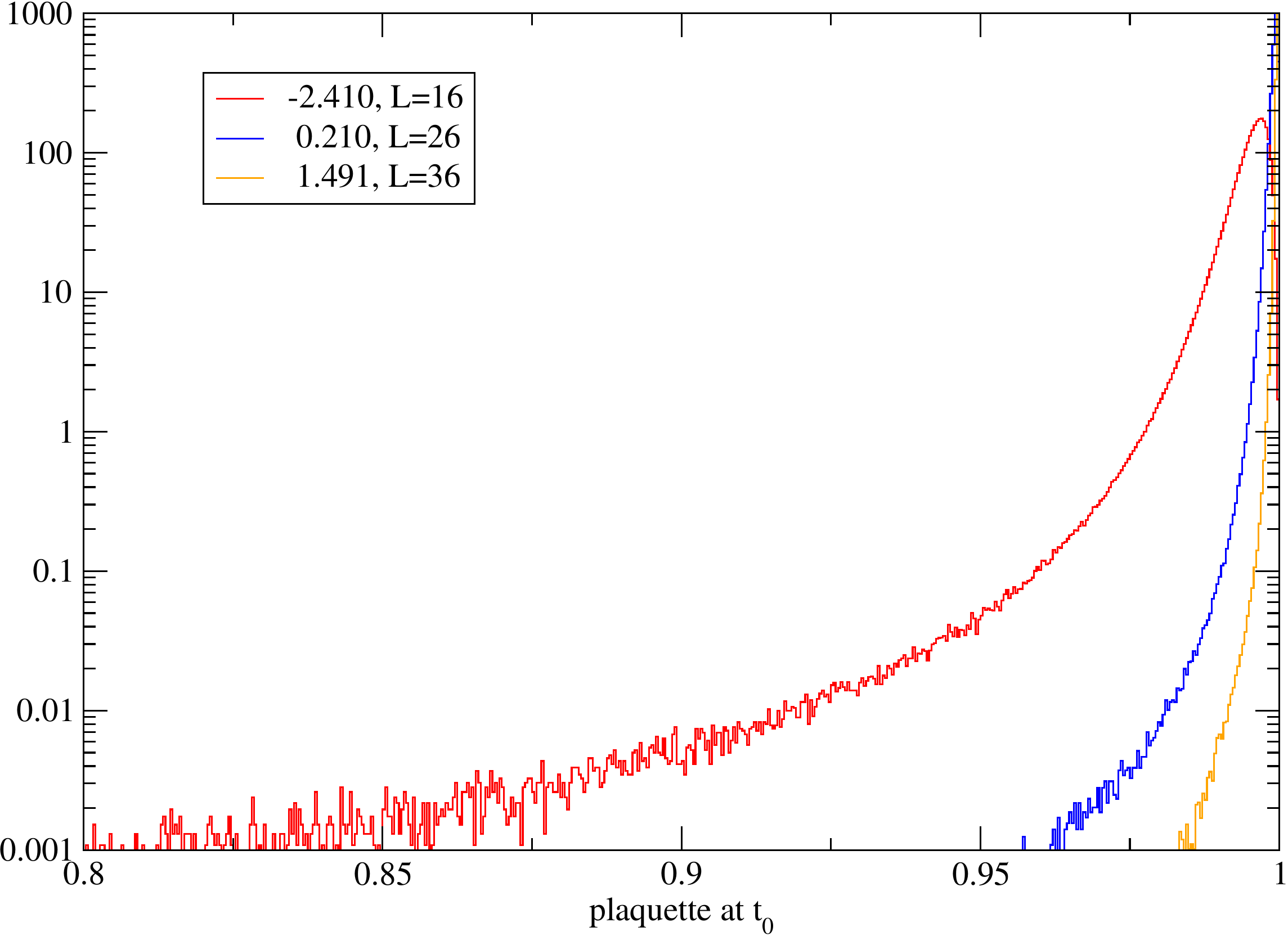}
  \caption{Distribution of the plaquette values of the gauge field
    configurations at flow time $t_0$. The left panel corresponds to the
    Wilson ensembles while the right panel shows the improved action
    ensembles, both at flow time $t_0$.\label{fig:plaquette_distr_t0}}
\end{figure}

\begin{figure}[thb]
  \centering
  \includegraphics[width=0.49\textwidth]{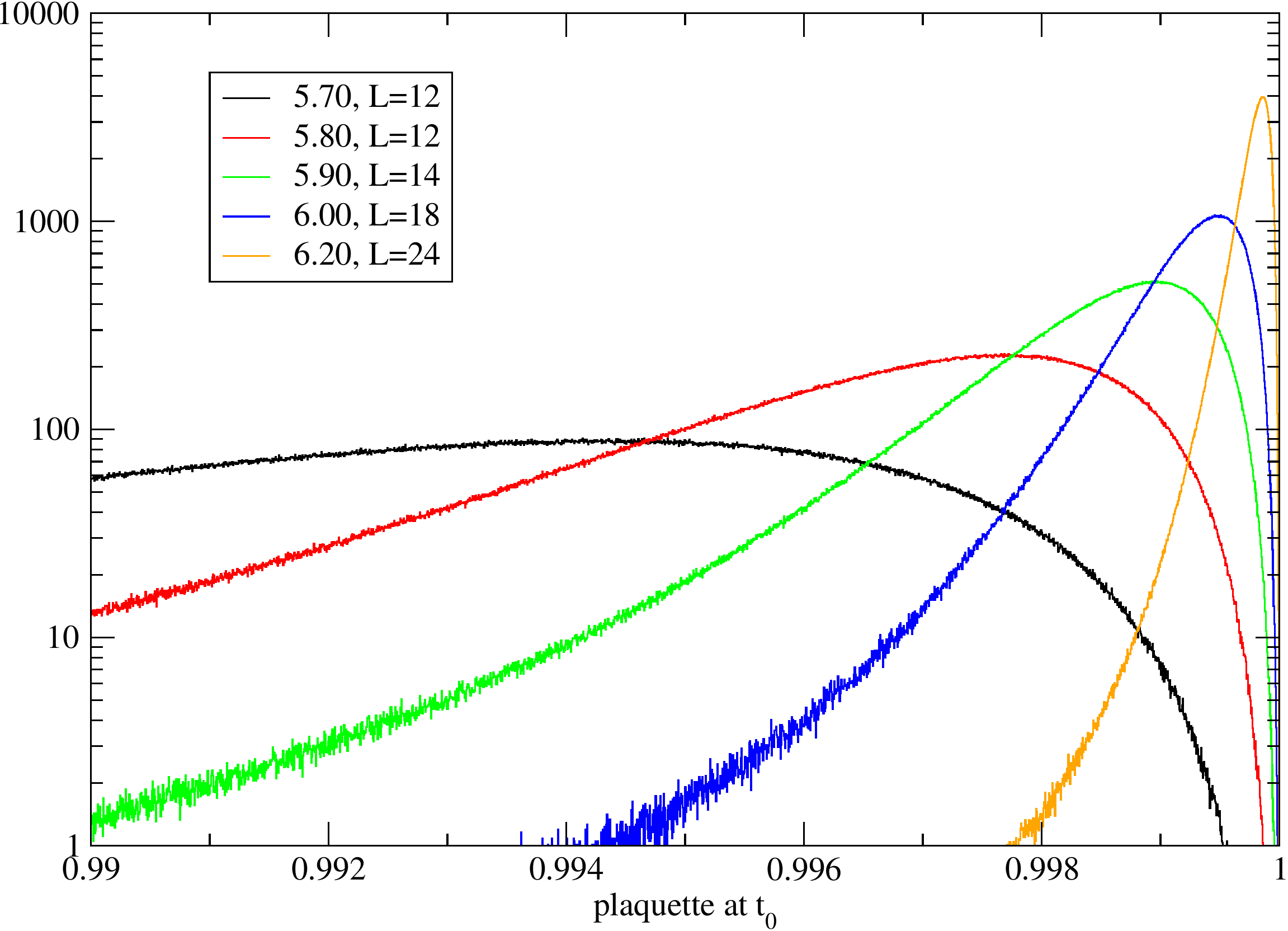}
  \includegraphics[width=0.49\textwidth]{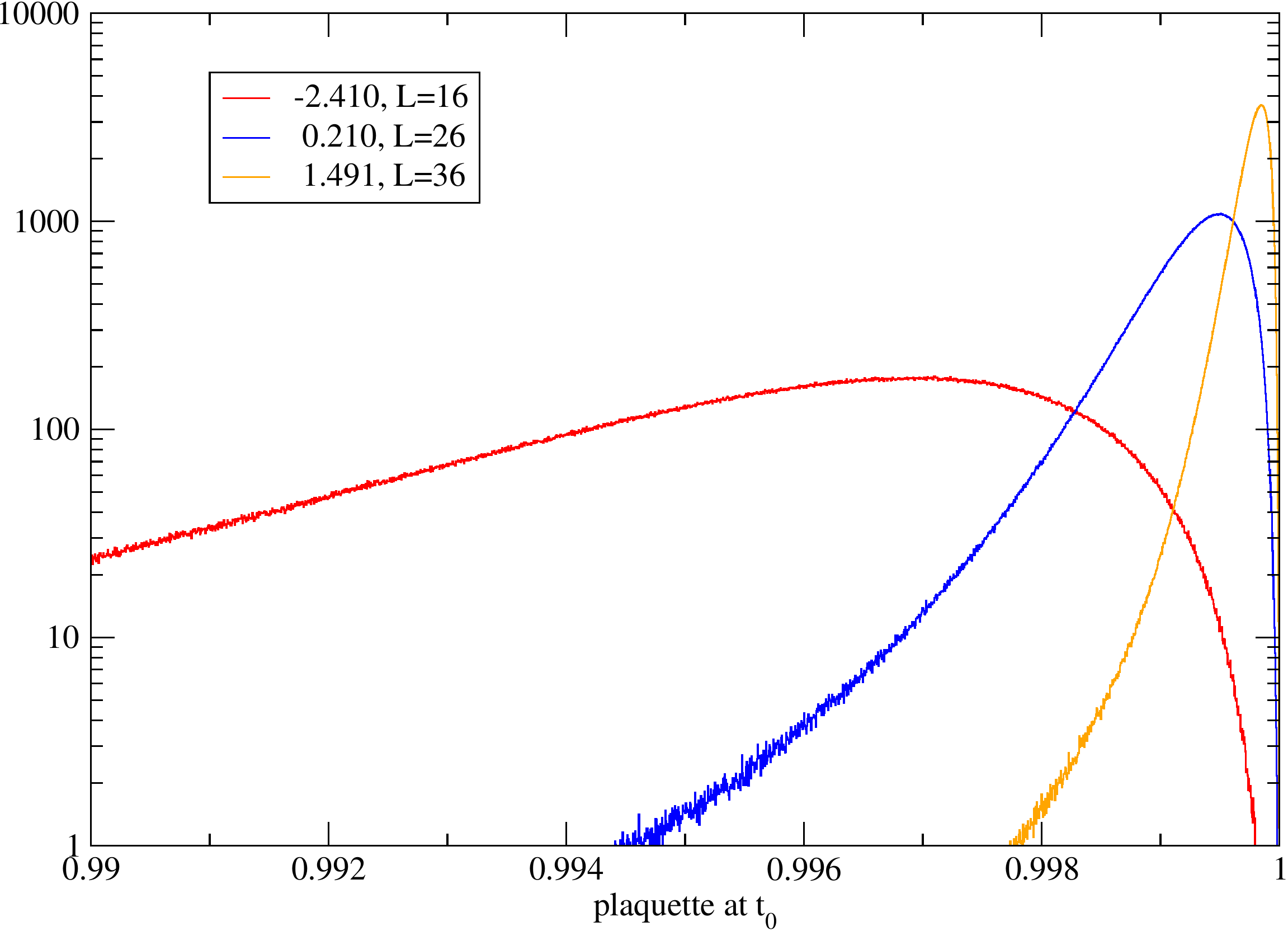}
  \caption{Magnification of the distribution of the plaquette values of the
    gauge field configurations. Left plot for Wilson ensembles, right plot for
    improved action ensembles at flow time
    $t_0$.\label{fig:plaquette_distr_t0_zoom}}
\end{figure}

To investigate further the origin of cut-off effects in observables related to
the gradient flow, we studied the distribution of the plaquette and its
evolution along the flow. This is shown in
figures~\ref{fig:plaquette_distr_orig}-\ref{fig:plaquette_distr_t0_zoom}. The
plaquette distribution on the original gauge configurations generated during
the Monte Carlo simulation are very different, by design: the new action
essentially suppresses large fluctuations, with the peak occurring further
from the limiting value where the plaquette is unity. As we tune to finer
lattice spacing, the distributions move smoothly. If we inspect the plaquette
distributions at flow time $t_0$ in figures
\ref{fig:plaquette_distr_t0} and \ref{fig:plaquette_distr_t0_zoom}, 
a different picture emerges. The Wilson and improved action ensembles have 
very similar distributions once the coarse fluctuations are smoothened out, 
with a rapidly decreasing tail of plaquette values away from 1. 
Using higher resolution to probe the region closest to 1, we see again 
the strong similarity between the
ensembles. The gradient flow washes out lattice artifacts contained in the
original lattice action, which are replaced by discretization effects of the
flow scheme instead. Obviously, the matching of the distributions
  between the two actions is in line with matching the lattice
  spacings from the gradient flow, and this is the reason why the
  gradient flow observables from the two actions show rather similar
  lattice artifacts.


\section{Algorithms and Cost Estimates}
\label{sec:costs}

The obvious drawback to using an improved action in lattice simulations is the
increased numerical cost, both in construction and in the Monte Carlo
simulation, where smaller updates are needed for e.g.~the Metropolis algorithm
to have a reasonable acceptance rate. Hence an increased time is necessary in
reaching a given numerical accuracy for a fixed computational resource. This
has to be balanced against the expectation of being able to determine
continuum results using coarser lattices than for the standard action. The
motivation to examine a new action as parametrized in eq.~\eqref{impaction} is
that using only powers of the basic plaquette minimizes the numerical cost,
while the suppression of large plaquette fluctuations is achieved in a way
very different than for lattice actions constructed along the Symanzik
improvement program.

Let us consider the numerical slowdown. For one sweep of a lattice volume, the
CPU time taken for the improved action relative to the Wilson action is
$t_{\rm imp}/t_{\rm Wil} = 1.9$, both for SU(2) and SU(3) gauge theory. In
order to estimate the autocorrelation time, we compare the squared relative
error of the Polyakov loop correlator at distance $t=L$, normalized to the
same number of sweeps, $(\Delta C(t)/C(t))^2 \times n_{\rm sweep}$, which we
found for SU(3) is 1.4 times larger with the improved action than for the
Wilson action (for an acceptence rate $\sim 0.5$ for both actions). For a
different determination of the autocorrelation time, we also examined the
torelon mass error, also normalized to the same number of sweeps, $(\Delta
m)^2 \times n_{\rm sweep}$. For this quantity we find an increase by a factor
of 1.5 going from the standard to the new action, essentially the same
value. Hence to achieve a given statistical accuracy our new action is about 3
times more expensive in computer time than the Wilson action. However, the
reduction in cut-off effects more than compensates for the increased cost.  As
a rough estimate, taking into account that the $a^2$ cut-off effect is reduced
by a factor $\sim 5$, to reach the same cut-off error and statistical error one
gains in computer time a factor $5^3/3\approx 40$, assuming the CPU time
needed for an independent configuration grows like $(L/a)^6$ (lattice volume
$\times$ critical slowing down).  We can only speculate that for full QCD the
gain could be even larger.

We use two different algorithms for Monte Carlo simulations with the improved
action. One is a standard Metropolis update, where a trial gauge link is
generated by a rotation of the original gauge link, followed by an
accept/reject decision. The rotation has an adjustable parameter, which allows
one to tune to whatever desired acceptance rate. A different algorithm is
where a trial gauge link is generated using the standard heathbath algorithm
for the Wilson action, but with a bare coupling $\beta' \ne \beta$. The trial
gauge link is then accepted or rejected with a Metropolis step based on the
change of the action $(\beta-\beta') w + \gamma w^q$. The adjustable parameter
here is $\beta'$ which can be tuned to improve the acceptance rate, however
the rate cannot be made arbitrarily close to 100\%. We find similar efficiency
between the two algorithms in our Monte Carlo results.

\section{Conclusions}
\label{sec:conclusion}

The type of lattice action we propose and study in this paper is somewhat
unusual, given that it does not have the usual naive continuum limit. On the
basis of universality, the essential elements are the dimensionality of the
system and that the lattice action has the correct internal symmetries. Our
numerical results fully support this view and show beyond any doubt that our
chosen discretization gives the correct continuum theory.

The findings regarding suppression of lattice artifacts depend on the
observable in question. In the case of spectral quantities such as the torelon
masses it was possible to almost fully remove cut-off dependence on the
coarsest lattice we simulate. For non-spectral quantities such as the static
quark potential and force, and observables given by the gradient flow such as
the lattice scales $t_0$ and $w_0$, improvement in the operators is necessary
beyond just improvement of the action, as can be seen from the reduction but
not the removal of artifacts. We did not measure other spectral quantities
such as the glueball spectrum, the critical temperature $T_c$ and the string
tension $\sigma$, since those studies would require larger numerical
simulations. The tuning strategy we choose is based on the torelon
spectrum. The parametrization is general -- one could extend it without
increasing the numerical cost by adding an adjoint plaquette term to the
action, which opens up the possibility that additional tuning of the extra
parameters would give even further reduction of lattice artifacts.

Our study of the gradient flow and its use in scale setting, as well as
observables given by higher order derivatives of the renormalized action
density, is a useful test of the accuracy of the scheme and of the systematics
due to lattice artifacts, as well as a test for universality.

The newly proposed lattice action is cheap and would be straightforward to
incorporate into simulations with dynamical fermions. The crucial question is
how much of the improvement carries over to such simulations, which remains to
be investigated.

\section{Acknowledgments}

The research leading to these results has received funding from the
Schweizerischer Nationalfonds and from the European Research Council
under the European Union's Seventh Framework Programme
 (FP7/2007-2013)/ ERC grant agreement 339220. We also acknowledge support by 
the US National Science Foundation under the grants NSF 0970137 and 1318220. 
KH wishes to thank the Institute for
Theoretical Physics and the Albert Einstein Center for Fundamental Physics at
the University of Bern for their support.


\end{document}